\documentclass[twocolappendix]{emulateapj-rtx4}
\usepackage{amsbsy,amsmath,bm,amssymb}
\usepackage{graphicx}
\usepackage{epstopdf}
\usepackage{natbib}

\newcommand{\Om}{\Omega_m}

\newcommand{\hmpc}{h^{-1}{\rm Mpc}}
\newcommand{\hmsun}{h^{-1}M_\odot}
\newcommand{\cm}{{\rm cm}}

\newcommand{\beqn}{\begin{eqnarray}}
\newcommand{\eeqn}{\end{eqnarray}}

\newcommand{\ben}{\begin{enumerate}}
\newcommand{\een}{\end{enumerate}}
\newcommand{\bit}{\begin{itemize}}
\newcommand{\eit}{\end{itemize}}
\newcommand{\beq}{\begin{equation}}
\newcommand{\eeq}{\end{equation}}

\newcommand{\lp}{\left(}
\newcommand{\rp}{\right)}

\def\f#1#2{\frac{#1}{#2}}

\newcommand{\HI}{H{\sc ~i}}

\newcommand{\HeI}{He{\sc ~i}}
\newcommand{\HeII}{He{\sc ~ii}}

\newcommand{\Msun}{{M_{\odot}}}

\newcommand{\zcol}{z_{\rm coll}}

\begin{document}

\title{A Physical Understanding of how Reionization Suppresses Accretion onto Dwarf Halos}
\author{Yookyung Noh\altaffilmark{1}, Matthew McQuinn\altaffilmark{1}$^{,}$\altaffilmark{2}}
\altaffiltext{1} {Department of Astronomy, University of California, Berkeley, CA 94720, USA\\}
 \altaffiltext{2} {Hubble Fellow; mmcquinn@berkeley.edu\\}	 
      
\begin{abstract}
We develop and test with cosmological simulations a physically motivated theory for how the interplay between gravity, pressure, cooling, and self-shielding set the redshift--dependent mass scale at which halos can accrete intergalactic gas. This theory provides a physical explanation for the halo mass scale that can accrete unshocked intergalactic gas, which has been explained with ad hoc criteria tuned to reproduce the results of a few simulations.  Furthermore, it provides an intuitive explanation for how this mass scale depends on the reionization redshift, the amplitude of the ionizing background, and the redshift.  We show that accretion is inhibited onto more massive halos than had been thought because previous studies had focused on the gas fraction of halos rather than the instantaneous mass that can accrete gas.   A halo as massive as $10^{11}\Msun$ cannot accrete intergalactic gas at $z=0$, even though typically its progenitors were able to accrete gas at higher redshifts.  We describe a simple algorithm that can be implemented in semi-analytic models, and we compare the predictions of this algorithm to numerical simulations. 
\end{abstract}


\section{introduction}
How gas travels from the intergalactic medium into halos and ultimately onto galaxies is an important input for models of galaxy formation.  It is thought that gas accretion is inhibited at halo masses of less than $10^{9}-10^{10}~\Msun$, for which thermal pressure from the photoionized intergalactic medium (IGM) suppresses accretion \citep{shapiro94, quinn96, bullock00, gnedin00, hoeft06, okamoto08}.  This mass scale is set by the complicated interplay between heating, cooling, self-shielding, and gravitational collapse processes.

The halo mass above which intergalactic gas can accrete is a necessary condition on which halos host galaxies, possibly setting the mass scale of the Milky Way's ultra-faint dwarf satellites.  
A common picture for the formation of these satellites is that they were born prior to reionization, when the lower Jeans' mass of intergalactic gas compared to later times allowed $\lesssim 10^9\Msun$ gas clouds to collapse \citep{bullock00, gnedin00, somerville02, benson02, busha10, lunnan12}. 
Recent measurements of the stellar ages of the ultra-faint dwarfs, which find advanced ages of $\approx 13~$Gyr, have added credence to this simple picture \citep{brown12}.  
A competing theory is that stellar feedback rather than reionization prevents the formation of low mass galaxies (e.g., \citealt{dekel03, mashchenko08, pontzen12}).  
Both reionization and stellar feedback play some role in shaping the properties of the smallest
galaxies that form in cosmological simulations (e.g., \citealt{pawlik09b, okamoto09, okamoto10, finlator11}).  In addition, ram pressure stripping by ambient hot gas in the Milky Way halo is a third process that suppresses star formation, at least in satellite galaxies (e.g., \citealt{penarrubia08}).

In order to disentangle the impact of reionization from other feedback processes, this paper develops a physical understanding for how reionization suppresses galaxy formation.
Here `reionization' refers to the same feedback process that other papers have termed the `photoionizing background':  The heating that pressurized the intergalactic gas occurred predominantly at reionization.  
Reionization resulted in a pervasive, largely homogeneous hydrogen--ionizing background that maintained intergalactic temperatures of $\sim 10^4$K and, hence, intergalactic Jeans' masses of $10^{9}-10^{11}~\Msun$.
We attempt to understand the physics of accretion onto a halo after reionization, as a function of the halo's redshift, the timing of reionization, and the amplitude of the photoionizing background.

Previous analytic and semi-analytic models parametrized the feedback from reionization with simple prescriptions.  In particular, many of these models assumed that the Jeans' mass (or similarly the `filtering mass'; \citealt{shapiro94, gnedin98, gnedin00}) evaluated at the mean density of the Universe determines the mass scale that is able to accrete \citep{busha10, lunnan12}.  Unsurprisingly, cosmological simulations show that the Jeans'/filtering mass is \emph{not} a good approximation.  \citet{hoeft06} and \citet{okamoto08}  adopted the criterion that gas can accrete if the halos equilibrium temperature (at which photoionization heating balances cooling) evaluated at an overdensity of $60-1000$ is greater than the halo virial temperature.  While this criterion more successfully reproduces the mass at which halos will retain more than half of their baryons \citep{hoeft06, okamoto08}, it is unclear physically why it should work.  

The impact of reionization on dwarf galaxies has also been studied with $1$D codes that follow the collapse of spherically symmetric perturbations \citep{thoul96, dijkstra04, sobacchi13a}.  These $1$D investigations quantified how the suppression of gas accretion onto dwarf galaxies depends on a variety of factors, such as the redshift, the thermal history, and the photoionizing background.  This paper provides an intuitive picture for many of the trends observed in the $1$D calculations. 

This paper is organized as follows.  Section \ref{sec:char_scales} summarizes the characteristic scales in the problem, using these scales to motivate a simple model for whether a gas cloud is able to collapse.  Section \ref{sec:simulations} describes our cosmological simulations, which are then used to test our model in Section \ref{sec:results}.  Finally, Section \ref{sec:merger_tree} implements our model in a halo merger tree.  This study assumes a flat $\Lambda$CDM cosmological model with $\Omega_m=0.27$, $\Omega_\Lambda=0.73$, $h=0.71$, $\sigma_8= 0.8$, $n_s=0.96$, $Y_{\rm He} = 0.24$, and $\Omega_b = 0.046$, consistent with the favored cosmology by the WMAP CMB experiment plus other large scale structure measurements \citep{larson11}.

\section{characteristic scales}
\label{sec:char_scales}
Many of the characteristic scales that factor into whether a halo is able to accrete surrounding gas can be expressed as functions of the temperature and density of ambient gas: (1) the cosmic mean density [and turnaround/virialization densities], (2) the Jeans' mass, (3) the densities and temperatures at which the cooling time equals the dynamical time, and (4) the density of gas that self-shields to ionizing photons.  In what follows, we define these scales and then use them to motivate a physical picture for how gas accretion is inhibited by reionization. This picture is based on the trajectories of gas parcels in the temperature--hydrogen number density ($T$-$n_H$) plane, where $n_H$ includes both atomic and ionic species.  

\subsection{cosmological density scales}
\label{ss:densities}
The cosmic mean hydrogen number density is
\begin{equation}
\langle n_H \rangle = 1.3\times10^{-5}\,\left( \frac{1+z}{4}\right)^3 ~~{\rm cm}^{-3},
\end{equation}
where $\langle ... \rangle$ indicates a volume average.
It is also useful to define the gas overdensity as $\delta_b \equiv n_H/\langle n_H \rangle - 1$.  Densities of relevance to our discussion, in addition to the mean ($\delta_b=0$), are the turnaround density ($\delta_b \approx 4.6$) and the virialization density ($\delta_b \approx 180$).  Turnaround -- when a region fully decouples from the Hubble flow such that its density starts increasing -- occurs at a redshift of $z_{\rm ta} = 2^{2/3}(\zcol +1) -1$, where $\zcol$ is the collapse (i.e., virialization) redshift.  These characteristic overdensities and redshifts are calculated assuming spherical collapse of a top hat density perturbation \citep{gunn72} and $\Omega_m(z) = 1$.  The latter applies in the assumed cosmology at redshifts of $z\gtrsim1$.

\subsection{the Jeans' mass}
\label{ss:Jeans}
The Jeans' mass -- the mass scale that can overcome pressure and collapse gravitationally -- is given by (e.g., \citealt{binney87})
\begin{eqnarray}
M_{J}&=& \f{4\pi}{3} \, \rho_m \lp \f {\pi} {k_{J}}\rp^{3},\\
 &=& 4\times 10^9 \left(\frac{T}{10^4\,{\rm K}} \right)^{3/2} \left(\frac{n_H}{10^{-3} \,{\rm cm}^{-3}} \right)^{-1/2}\Msun,
\label{eq:mJ}
\end{eqnarray}
where
\begin{equation}
 k_{J} \equiv c_s^{-1} t_{\rm dyn}^{-1}, ~~~~~  t_{\rm dyn} \equiv \lp4\pi G \rho_m \rp^{-1/2}.\\
\end{equation}   
Here, $\rho_m$ is the mass density of gas plus dark matter, $c_s$ is the sound speed (evaluated in eqn.~\ref{eq:mJ} for isothermal gas of primordial composition and adiabatic index $\gamma =5/3$), and $t_{\rm dyn}$ is the dynamical time.  

The coefficient of $4\times 10^9$ in equation~(\ref{eq:mJ}) is evaluated at ${n_H} = 10^{-3}\, {\rm cm}^{-3}$, which is roughly the density where the Jeans' criterion is most relevant in our models for collapse at $3 <z <6$ (\S\ref{ss:example}).  Smaller $n_H$ are relevant for halos at lower redshifts.  This coefficient also took $T=10^4$K, which is likely a good approximation after the reionization of hydrogen.   Reionization heated the intergalactic medium (IGM)
to $\sim 2 \times 10^{4}$~K from $\sim 10-1000~$K \citep{miralda94, mcquinn-Xray}.  Afterward, the gas in the Hubble flow cools with the expansion of the Universe, but it can only cool to $0.5-1\times10^4~$K as at these temperatures adiabatic cooling comes into balance with photo-heating \citep{hui97}.

While the Jeans' mass is derived by analyzing the growth of modes in a homogeneous medium, it is also the mass of a region that has diameter set by the distance a sound wave travels in a dynamical time.  In addition, to factors of order unity it sets when the thermal energy of a cloud equals the gravitational energy (ignoring the kinetic inertia of collapse, which is much smaller unless the cloud is near the virialization density).  Thus, even for the case of interest -- a collapsing gas cloud -- the Jeans' criterion approximates the mass scale at which pressure is able to respond to and halt collapse.  Overdense \HI\ absorbers in cosmological hydrodynamic simulations have been found to have sizes of roughly the Jeans' length \citep{schaye01, mcquinn-LL, altay11}.

\subsection{the equilibrium temperature where atomic cooling balances photoheating}

For primordial gas exposed to the extragalactic ionizing background, atomic transitions of \HI\ are the dominant coolant at $T \lesssim 5\times 10^4~$K.  At higher temperatures, \HeII\ cooling can dominate. The dominant heating process for collapsing gas is \HI\ and \HeI\ photoionization because the \HeII\ in collapsing regions tends to not be exposed to a significant ionizing background (see \S \ref{ss:selfshielding}).   The equilibrium temperature at which heating balances cooling unfortunately cannot be encapsulated with a clean analytic formula.  Table~\ref{table1} gives the equilibrium temperature of primordial gas exposed to a photoionizing background with $\Gamma_{-12}=1$ and $\Gamma_{-12}=0.1$, where $\Gamma_{-12}$ is the \HI\ photoionization rate in units of $10^{-12}~$s$^{-1}$.  We assume throughout the spectral index of the specific intensity [erg~s$^{-1}$Hz$^{-1}$sr$^{-1}$] is equal to $0$ and that the \HeI\ photoionization rate equals that of \HI, which roughly approximates what is found in ionizing background models \citep{haardt12}.  At $2<z<5$, observations find $\Gamma_{-12}=1$ \citep[e.g.,][]{becker13}, with smaller values estimated at lower and higher redshifts \citep{fan06, haardt12}.  In addition, we find that the equilibrium temperature is almost identical to the temperature at which the cooling time equals the dynamical time -- which is the more applicable criterion.  This equivalence results because of the exponential temperature dependence of collisional cooling.

\begin{table}
\caption{The equilibrium temperature at which photoheating balances cooling,$^{*}$ computed for the specified \HI\ and \HeI\ photoionization rates and a \HeII\ photoionization rate of zero.}
\begin{center}
\begin{tabular}{l l l}
&  $\Gamma_{-12} = 1$ &  $\Gamma_{-12} = 0.1$\\
\hline
$n_H$ [cm$^{-3}$] &  $T_{\rm eq}$ [$10^4\,$K] & $T_{\rm eq}$~[$10^4\,$K] \\
\hline
\hline
$1\times10^{-5}$& 41 &41\\
$3\times10^{-5}$& 22 &22\\
$1\times10^{-4}$& 8.5 &8.5\\
$3\times10^{-4}$& 5.9 &5.3\\
$1\times10^{-3}$& 3.1 &1.9\\
$3\times10^{-3}$& 2.3 & 1.6\\
$1\times10^{-2}$& 1.8 & 1.3 \\
$3\times10^{-2}$& 1.5 & 1.2\\
\end{tabular}
\end{center}
$^{*}$ To the quoted precision, this temperature also equals the temperature at which the cooling time is equal to the dynamical time.
\label{table1}
\end{table}

\subsection{the density at which gas self-shields} 
\label{ss:selfshielding}

For a gas cloud to self-shield to \HI-ionizing photons (i.e., has optical depth $\geq 1$ to $1~$Ry) requires \HI\ column densities of $N_{\rm HI} \geq \sigma_{\rm HI}^{-1} = 1.6\times10^{17}~$cm$^{-2}$, where $\sigma_{\rm HI}$ is the \HI\ photoionization cross section at $1~$Ry.  We denote the fraction of hydrogen that is in \HI\ as $x_{\rm HI}$, the CASE A recombination coefficient as $\alpha$, and the electron density as $n_e$ (which equals $n_H$ for ionized gas).  If the sizes of intergalactic absorption systems are set by the Jeans' length as previously argued and if the gas is in photoionization equilibrium with the background value such that $\Gamma x_{\rm HI} = \alpha \, n_e$, then the critical density to have an optical depth of unity at $1~$Ry is \citep{schaye01}
\begin{equation}
n_H \approx  0.004 \,{\cm^{-3}}\,\Gamma_{-12}^{2/3}   \left(\frac{T}{10^4K} \right)^{0.17}.
\label{eqn:schayer0}
\end{equation}

While equation~(\ref{eqn:schayer0}) gives the density that starts to self-shield to an ionizing background, it is not the density that is able to self-shield sufficiently and stay neutral.  Studies have found that hydrogen columns of $10\, \sigma_{\rm HI}^{-1}$ are required to self shield sufficiently to remain neutral \citep[in part because higher energy photons have smaller optical depths]{altay11, mcquinn-LL}, requiring densities above
\begin{equation}
n_H \approx  0.02 \,{\cm^{-3}}\,\Gamma_{-12}^{2/3}   \left(\frac{T}{10^4K} \right)^{0.17}.
\label{eqn:SS}
\end{equation}
This number agrees with the radiative transfer calculations of \citet{faucher10} with $\Gamma_{-12}\approx 0.5$, which find $n_H \approx 0.01\,{\cm^{-3}}$ reliably describes the transition to neutral gas.

Capturing self-shielding is important for modeling accretion onto halos for two reasons:  First, both atomic cooling and also the contribution of local sources of radiation to the background (which we have ignored) become more significant in self-shielding regions once equation~(\ref{eqn:schayer0}) is satisfied.\footnote{One can show that for a typical region, the internal production of ionizing photons can \emph{only} have a significant impact on the photoionization rate in locations that can self-shield \citep{miralda05, rahmati13}.}  Second, collapsing regions that satisfy equation~(\ref{eqn:SS}) over their entire history are never photoheated.  We can estimate the redshifts when the latter occurs using the spherical
collapse model:  A region is always above this critical density for being fully self-shielded if it collapses at
\begin{equation}
\zcol > 9.3\left(\frac{\Gamma_{-12}}{0.1}\right)^{2/9} - 1,
\end{equation}
where we have equated the turnaround density to the density that self-shields as given by equation~(\ref{eqn:SS}) with $T= 10^4$K.  As previously mentioned, $\Gamma_{-12} \approx 0.1$ is consistent with observational estimates at 
 $z\approx 6$ \citep{fan06, calverley11}, and the average background intensity is almost certainly 
smaller with increasing redshift.  Thus, galaxies that are formed from collapse at $z\gtrsim8$ are typically fully self-shielded and not impacted by photoionization feedback.

\HeII\ self-shields more easily than \HI.   At $z=2.5$ -- when quasars peak in abundance -- simple estimates show that the \HeII\ self-shields at $\sim 30$ times lower densities than the \HI\ \citep{mcquinn13}.  At lower and higher redshifts, the \HeII-ionizing background in most regions is expected to be weaker and, hence, the critical density at which self-shielding occurs lower \citep{worseck11}.  Thus, collapsing regions are likely to be self-shielded to $4\;$Ry photons once they reach densities where cooling is important, which justifies setting the \HeII\ photoionization rate equal to zero (as is done subsequently).   A more significant \HeII\ ionizing background would not alter our picture for most redshifts as it would inhibit cooling only in halos collapsing at $z \lesssim 1$ (\S\ref{ss:example}).

\subsection{gas particle trajectories in the $T-n_H$ plane}
\label{ss:example}

\begin{figure*}
\begin{center}
\includegraphics[width=7.in]{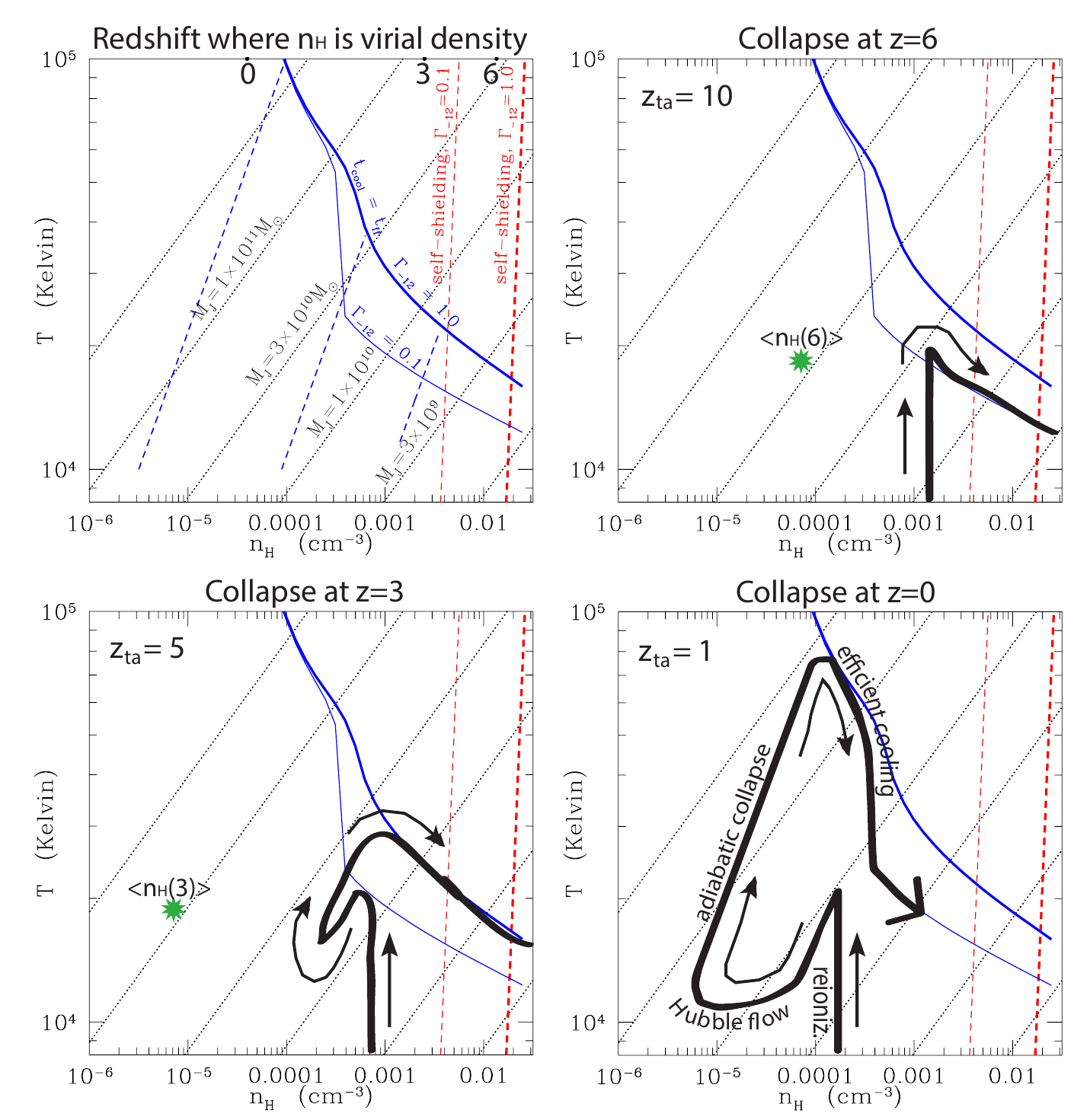} 
\end{center}
   \caption{Illustration of how gravitationally unstable gas clouds with the specified collapse redshifts travel in the $T-n_H$ plane.  The black dotted diagonal lines represent contours of constant Jeans' mass. The blue solid curves show the equilibrium temperature.  The nearly vertical, red dashed lines are the thresholds at which gas is fully self-shielded (assuming the $N_{\rm HI}=10 \,\sigma_{\rm HI}^{-1}$ criterion for self-shielding).  The two equilibrium temperature curves as well as the two self-shielding curves are computed for $\Gamma_{-12} = 0.1$ (thin) and $1$ (thick).  In the top-left panel, the three blue dashed diagonal lines that intersect the cooling curves are adiabats (i.e., $T \propto n_H^{2/3}$).  The thick black solid curves in the other three panels illustrate schematic trajectories of gas particles accreted onto halos at $z=6$, $z=3$, and $z=0$.  The main phases of the trajectory are labeled in the $z=0$ panel.  The green stars in the $z=6$ and $z=3$ panels are the cosmic mean hydrogen density at the collapse redshift -- the density at which many studies had evaluated the Jeans' mass (or filtering mass) to determine the minimum halo mass that can host a galaxy.}
   \label{fig:theorycurve}
\end{figure*}

The top-left panel in Figure \ref{fig:theorycurve} shows where the three characteristic curves described in \S2.2--2.4 (as well as curves representing different adiabats) lie in the $T-n_H$ plane: 
\begin{enumerate} 
\item The black dotted diagonal lines represent contours of constant Jeans' mass. 
\item The two blue solid curves show where the cooling rate balances the photoheating rate, $T_{\rm eq}(n_H)$, for the cases $\Gamma_{-12} = 0.1$ (thin) and $1$ (thick).
\item The two red (nearly vertical) dashed lines are the thresholds at which gas is fully self-shielded (and hence neutral) for $\Gamma_{-12} = 0.1$ (thin) and $1$ (thick).  These curves use the $N_{\rm HI}=10 \sigma_{\rm HI}^{-1}$ criterion for self-shielding.  Partial self shielding such that $N_{\rm HI}=\sigma_{\rm HI}^{-1}$ occurs at $10^{2/3}$ smaller densities.
\item The blue dashed lines show three example adiabats ($T \propto n_H^{2/3}$).  
\end{enumerate}
An expanding or collapsing cloud will approximately travel along an adiabat in the region where (1) its cooling time is longer than the dynamical time (i.e., below the solid blue curves) and (2) its temperature is greater than $10^4$K, below which photo-heating becomes important \citealt{hui97}).  Finally, the top axis in the top-left panel shows the virialization density for collapse at $z=0,~3$, and $6$. 

The thick solid curve in each of the other three panels of Figure~\ref{fig:theorycurve} shows a schematic trajectory for a gravitationally unstable gas cloud that collapses at $z=6$, $z=3$, and $z=0$, assuming that the overdensity follows that expected in the spherical collapse model and that reionization heated the gas to $2\times 10^4\;$K at $z=9$.  This is the anticipated temperature if stars reionized the Universe \citep{miralda94, mcquinn-Xray}.  In addition, select curves in the top-left panel appear in the three other panels. 

Let us start with the thick solid trajectory shown in the top-right panel in Figure \ref{fig:theorycurve} for which $\zcol = 6$.  This trajectory's turnaround redshift is approximately its reionization redshift.  Thus, at turnaround the gas is heated by reionization to values that are sufficient to cool, radiate away energy, and continue collapsing, following the $T_{\rm eq}(n_H)$ curve to higher densities.
If instead the cloud collapses onto a halo at lower redshifts than $\zcol = 6$ (the bottom two panels), it would first cool adiabatically at it expands in the Hubble flow, until it reaches a floor in the temperature at $\approx 10^4$K when cooling balances photo-heating. 
At turnaround, it would begin to collapse and heat up adiabatically (unless it were shock heated during this phase, which would cause it to ascend to a higher adiabat).
Finally, once a gas parcel collapses to densities at which atomic cooling becomes important, the gas is able to radiate away energy and, hence, follow the blue cooling curves to the right.  In Figure~\ref{fig:theorycurve}, the gas parcels that are accreted at $z=6$ and $z=0$ follow the $\Gamma_{-12}=0.1$ cooling curve once cooling becomes important, while the gas that was accreted onto a halo at $z=3$ follows the $\Gamma_{-12}=1$ curve.  These choices are motivated by observational estimates of $\Gamma_{-12}$ \citep{bolton05}.\footnote{Gas could become heated at the virial shock to higher temperatures than in our illustrative trajectories, especially if it is collapsing onto a $>10^{12}\Msun$ halo \citep{keres05}. However, in all but the most massive halos in the Universe, such shocked gas is able to cool in much less than the Hubble time and continue condensing.}

In our picture, the trajectories shown in Figure~\ref{fig:theorycurve} would represent gas collapsing into a dark matter halo whose mass is greater than the Jeans' masses the trajectory intersects.
However, the collapse would halt if (roughly) the mass of the accreting halo is less than the Jeans' mass of the gas at any point along the trajectory (meaning that gravity cannot overcome pressure).   
Notice that the lower the redshift that the gas is accreted, the larger the halo mass that is required to overcome pressure (i.e., the Jeans' mass, which is constant along the dotted curves, increases moving from the bottom right towards the top left in each panel). 
Thus, whether a gas parcel can be accreted depends on whether its halo is massive enough so that it is Jeans' unstable at all densities along its trajectory in the $n_H-T$ plane.  We test this simple model in the ensuing sections.  

\subsection{Previous models}
\label{ss:previous}
The picture for gas accretion given in \S\ref{ss:example} contrasts with previous models in the literature.  \citet{gnedin00} argued that the  `filtering mass', $M_F$ -- 
the expanding Universe analog of the Jeans' mass -- sets the scale at which gas can accrete at redshift $z$:
\begin{equation}
M_F(z) = \frac{4\pi}{3}\rho_b\left(\frac{\pi}{k_J}\right)^3 f(z, z_{\rm rei})^{3/2},
\label{eqn:Mf}
\end{equation}
where $z_{\rm rei}$ is the reionization redshift and
\begin{eqnarray}
	f(z, z_{\rm rei}) &=& \frac{3}{10}\left[1+4\left(\frac{1+z}{1+z_{\rm rei}}\right)^{2.5}-5\left(\frac{1+z}{1+z_{\rm rei}}\right)^{2}\right]. \nonumber
\end{eqnarray}
This formula for $M_F$ assumes that the temperature of $\delta_b=0$ gas after reionization is $10^4~$K \citep{gnedin98}, which approximates the thermal history in our simulations.  In detail, \citet{gnedin00} defined $M_F$ to be a factor of $8$ larger than the above, but later studies found that the above definition is more successful \citep[e.g.,][]{2009MNRAS.399..369N}.

In contrast, \citet{okamoto08} argued that only halos for which the equilibrium gas temperature at overdensity $\delta_*$ is less than the halo virial temperature can accrete gas.  This criterion is given by
\begin{equation}
	M_{acc}(z, \delta_*) = \f1{GH_{0}} \lp \f{2 k_{B} T_{eq}(\delta_*)}{\mu m_{p} (1+z)}  \rp^{3/2}
	\lp \f{\Delta_{c}(z) \Omega_{m, 0}}{2 \Om(z)} \rp^{1/2}.
	\label{eqn:Ma}
\end{equation}
where $\Delta_{c} (z) = 18\pi^{2} + 82d - 39 d^{2}$, and
$d = \Om(z) - 1 = \Omega_{m, 0} (1+z)^{3}/ (\Omega_{m, 0} (1+z)^{3} + 1 - \Omega_{m, 0}) - 1$.
\citet{okamoto08} assumed that the temperature of 
   the gas at the edge of a halo controls the accretion, setting $\delta_* = \delta_{\rm vir}/3$, where $\delta_{\rm vir}$ is the the halo virial overdensity.  

Lastly, \citet{hoeft06} provided the parametrization for the mass scale that retains half of the gas:
\begin{equation}
	\f{M_{c}(z)}{10^{10} \hmsun} = \lp \f{\tau(z)}{1+z} \rp^{3/2} \lp \f{\Delta_{c}(0)}{\Delta_{c}(z)} \rp^{1/2},
	\label{eqn:Mc}
\end{equation}
where $\tau(z) = 0.73 \times (1+z)^{0.18} \exp[-(0.25z)^{2.1}]$.  Equation~(\ref{eqn:Mc}) has a similar form to equation~(\ref{eqn:Ma}), and in fact is motivated with the same physical picture of the equilibrium temperature being equal to the virial temperature.  However, there are two major differences between the \citet{hoeft06} and \citet{okamoto08} models:  1) Equation~(\ref{eqn:Mc}) is most similar to equation~(\ref{eqn:Ma}) if it is evaluated at an overdensity of $10^3$ rather than $\delta_{\rm vir}/3$ \citep{hoeft06}, and 2) $M_c$ is the mass above which a halo retains at least half its gas rather than the mass that can instantaneous accrete.  In contrast, the \citet{okamoto08} model requires halo merger trees in addition to $M_{\rm acc}$ to calculate the mass at which halos retain half their gas.

  In what follows, we will compare our picture with these models.

\section{simulations}
\label{sec:simulations}

We aim to compare our picture for accretion outlined in the previous section with gas accretion in 3D cosmological hydrodynamic simulations.  We use the smooth particle hydrodynamics (SPH) code GADGET-3 \citep{springel01}, run in a mode where gas particles are turned into stars once $\delta_b>500$ in order to speed up the computation.  All of our simulations are in a 10~$\hmpc$ periodic box,  with either 256$^{3}$ or 512$^{3}$ SPH particles and an identical number of dark matter particles.   
These simulations were started at $z=100$ and initialized with 2$^{\rm nd}$ order Lagrangian perturbation theory applied to a glass particle distribution \citep{crocce06}, and all of our simulations are initialized with the same random numbers.  In the $512^3$ simulations, which are primarily for demonstrating convergence, additional random numbers were generated for the modes not in the other simulations.  The minimum halo mass studied (linking $32$ particles) in the $256^3$ and $512^3$ simulations is $2.0\times10^{8} M_{\odot}$ and $2.4\times10^{7} M_{\odot}$, respectively.\footnote{We select dark matter halos with GADGET-3's built in Friends-of-Friends halo finder with linking length set to $0.2$.}

In all of the simulations, snapshots were output on intervals of half of a dynamical time for gas with $\delta_b = 180$, resulting in $45$ snapshots between $1 < z <9$.  This frequency of outputs ensures that large changes in the density of $\delta_b\lesssim180$ gas are unlikely between adjacent snapshots.  The simulations were terminated at $z=1$ as the modes on the box scale become nonlinear at lower redshifts.

The simulations model reionization as an instantaneous process, ionizing the \HI\ and \HeI\ at $z=9$ and boosting the temperature to $1\times 10^{4}~$K.  \HeII\ reionization (which occurs at $z\sim 3$;  \citealt{mcquinn-HeII}) is ignored for simplicity.  After \HI\ reionization, gas particles are kept ionized by a photoionizing background.  In simulation SimG1, we use a photoionization rate for the \HI\ and \HeI\ of $\Gamma_{-12} = 1$.\footnote{Many previous studies used  $J_{21}$ -- the specific intensity intensity in units of $10^{-21}~{\rm erg~s^{-1}~Hz^{-1}~sr^{-1}}$ -- instead of $\Gamma_{-12}$.  $\Gamma_{-12} = 1$ for the \HI\ corresponds to $J_{21} = 0.25$ at the Lyman-limit for a spectral index of $0$.}  This photoionization rate is consistent with what is measured at $z=2-4$ \citep[e.g.,][]{bolton05}.  Simulation SimG10 instead uses $\Gamma_{-12} = 10$, which makes the hydrogen more ionized and hence suppresses cooling.  In addition, all of our simulations ignore self-shielding, except SimG1SS which allows gas particles to self-shield if they are more dense than the density criterion of equation~(\ref{eqn:SS}) evaluated with $T=10^4~$K.  All of our calculations that include an ionizing background take it to have a spectral index in specific intensity of zero.  (The spectral index only mildly impacts the \HI\ and \HeI\ photoheating rates.)  Our ionizing background model is simpler than in previous studies that used complex ionizing backgrounds a la \citet{haardt96}, but it captures the relevant physics in a more controlled manner.   Our simulations do not include stellar or active galactic nuclei feedback prescriptions.  
  
 Lastly, we ran corresponding $256^3$ and $512^3$ SPH particle adiabatic simulations (SimAd and SimAd512).  These simulations have neither global heating due to reionization nor cooling.  Because the adiabatic simulations have an unheated IGM (aside from structure formation shocks), we will use the differences between these simulations and the others to isolate the impact of gas pressure on gas accretion.

\begin{table}
\caption{The Gadget-3 simulation specifications, including the number of gas particles, $N_g$, the \HI\ and \HeI\ photoionization rate, $\Gamma_{-12}$, and the gas particle mass in units of $10^6\Msun$, $m_{\rm SPH}$.}
\begin{center}
\begin{tabular}{l c c c c}
Simulation & Box Size [Mpc$/h$] & $N_g$ & $\Gamma_{-12}$ & $m_{\rm SPH}$\\
\hline
\hline
SimAd  & $10$ & $256^3$ & 0 & 1.1 \\
SimG1  & $10$ & $256^3$ & 1 & 1.1 \\
SimG10 & $10$ & $256^3$ & 10 & 1.1 \\
SimG1SS & $10$ & $256^3$ & 1$^{*}$ & 1.1 \\
SimAdN512 & $10$  & $512^3$ & 0 &  0.13 \\
SimG1N512 & $10$  & $512^3$ & 1 &  0.13 \\
\end{tabular}
\end{center}
$^{*}$ except with $\Gamma_{-12}=0$ in self-shielding regions, using equation~(\ref{eqn:SS}) with $T=10^4$K for the critical density for self-shielding
\label{table:sim}
\end{table}

\section{gas accretion in cosmological simulations}
\label{sec:results}

To test our model with the simulations, we isolate gas particles  that would have been accreted onto halos if the Universe had never been heated by reionization.  
These particles are likely to be the ones that were accreted onto halos in our adiabatic simulations, where the intergalactic gas temperatures are generally very low.  Since all of our simulations start with the same initial conditions, we identify initially co-spatial particles in the other (non-adiabatic) simulations with the particles in the corresponding adiabatic simulation.  We selected gas particles that are accreted at $\zcol$ in the adiabatic simulations with the following three conditions:
\begin{enumerate}
\item fall within $1\, r_{\rm vir}$ of any halo's friends-of-friends center of mass,
\item have overdensity above $200$ at $z_{\rm coll}$,
\item have overdensity below $200$ in all of the higher redshifts snapshots ($z>z_{\rm coll}$),
\end{enumerate}
where $r_{\rm vir}$ is computed with the spherical collapse model.
  Following the corresponding gas particles in the simulations that include the heating from reionization (i.e., SimG1 and SimG10) tests how gas pressure impacts the particles' evolution.

\begin{figure}
\begin{center}
   \includegraphics[width=3.7in]{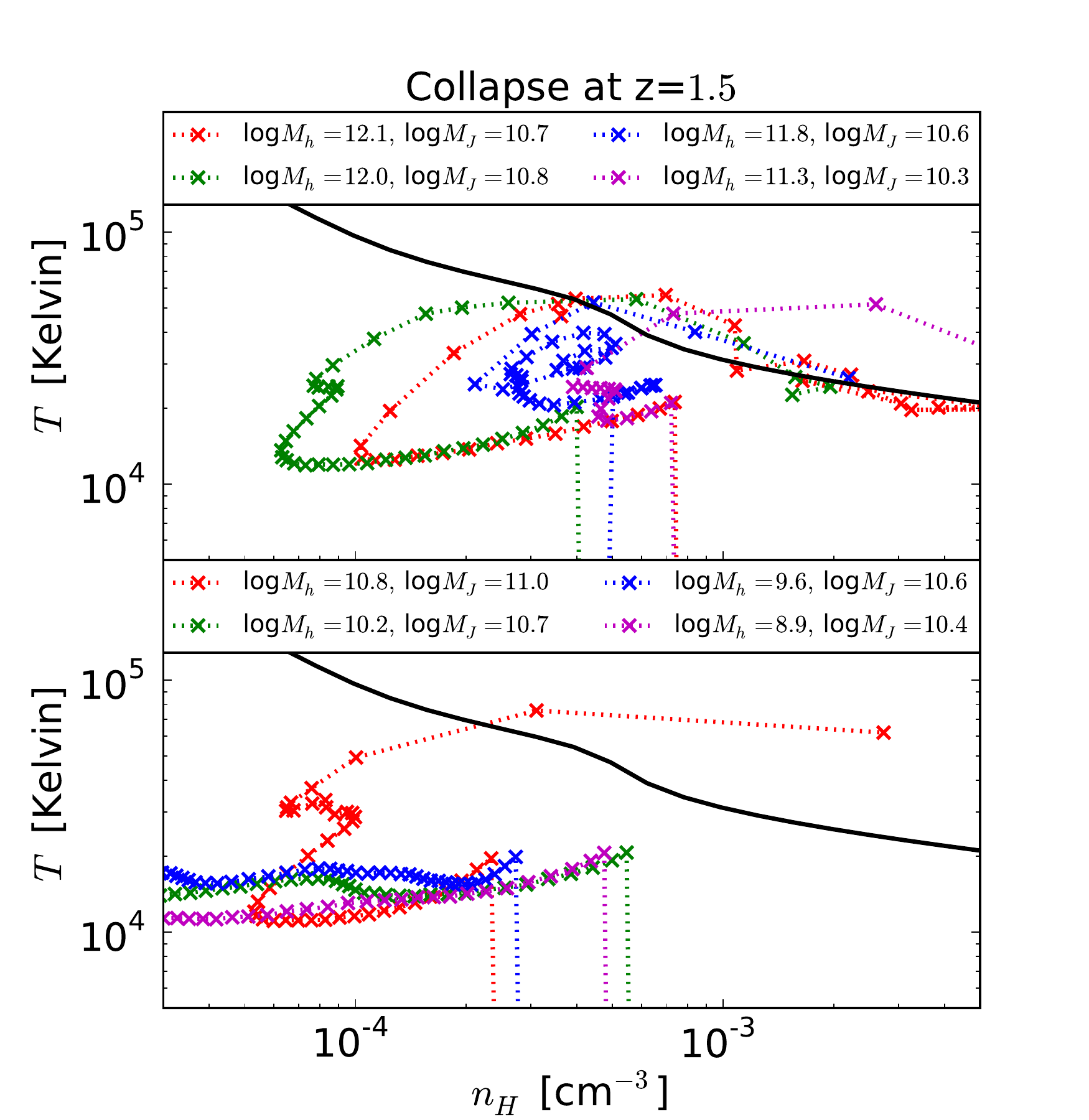}	
\end{center}
   \caption{The dotted curves in both panels show the trajectories of four gas particles in simulation SimG1.  The markers indicate the state of the gas at the times of the simulation outputs.  These SPH particles were selected so that their corresponding particle in simulation SimAd crossed the overdensity threshold of $\delta_b > 200$ at $z=1.5$ and never previously.  The vertical component of each trajectory owes to the simulation's instantaneous reionization at $z=9$, and the final point in each trajectory corresponds to $z=1.5$.   The solid curves are the equilibrium temperature at which photoheating balances atomic cooling.  The legend lists for each trajectory the halo mass onto which the corresponding particle was accreted in simulation SimAd 
as well as the trajectory's maximum Jeans' mass before crossing the cooling curve, $M_J^{\rm max}$.  \emph{Particles that have $M_h < M_J^{\rm max}$ are less likely to collapse to high densities, although we find in \S\ref{ss:simpic} that a more accurate criterion is $M_h < M_J^{\rm max}/4$.}
}
\label{fig:Tdsim}
\end{figure}

\begin{figure}
\begin{center}
  \includegraphics[width=3.7in]{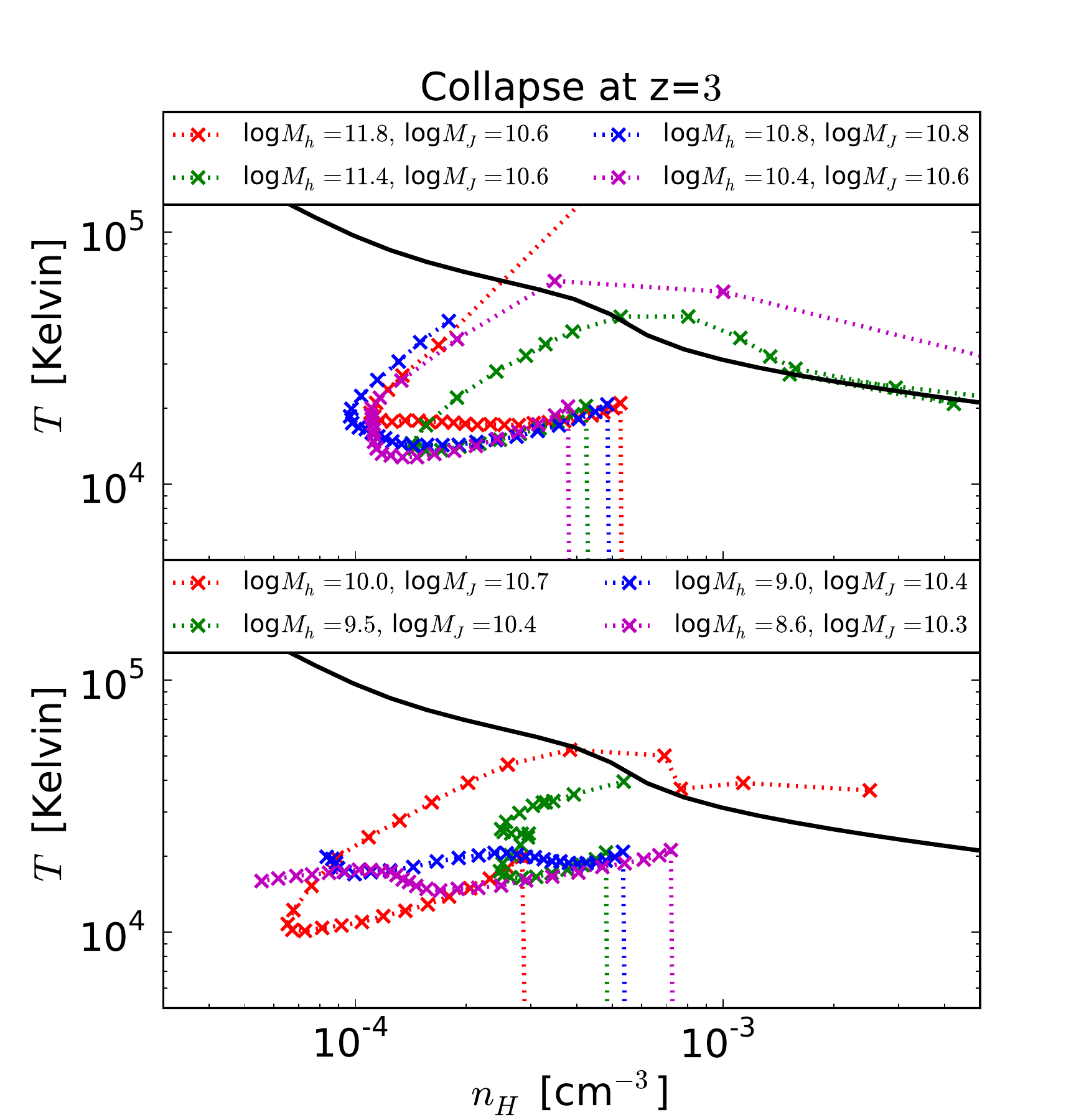}
\end{center}
   \caption{The same as Figure~\ref{fig:Tdsim} except the curves show trajectories of gas particles that would have been accreted at $z=3$ in the absence of gas pressure.
\label{fig:Tdsim2}}
\end{figure}

\begin{figure}
\begin{center}
 \includegraphics[width=3.7in]{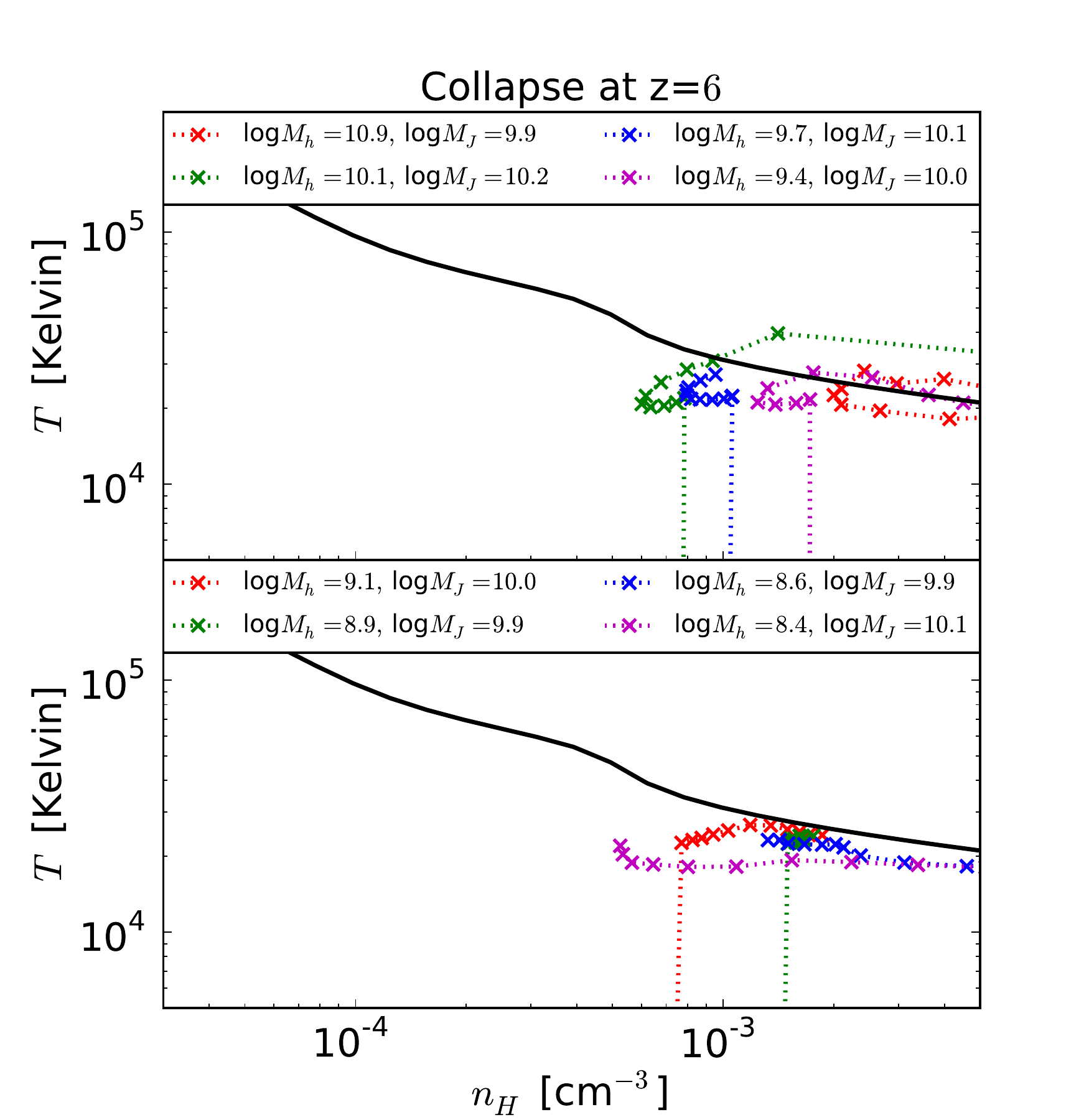}
\end{center}
   \caption{The same as Figure~\ref{fig:Tdsim} except the curves show trajectories of gas particles that would have been accreted at $z=6$ in the absence of gas pressure. 
\label{fig:Tdsim3}}
\end{figure}

It is illustrative to observe a few example $n_{H}-T$ trajectories of the selected gas particles.  The dotted curves in Figures~\ref{fig:Tdsim}-\ref{fig:Tdsim3} show such trajectories from SimG1 (with the markers indicating the times of simulation outputs).  For reference, the solid curves are $T_{\rm eq}$.
Figures~\ref{fig:Tdsim}, \ref{fig:Tdsim2}, and \ref{fig:Tdsim3} show particles with $z_{\rm coll}\approx1.5, \, 3$, and $6$, respectively.  
The two panels in each figure both show the trajectories of four gas particles, selected to span a range of halo masses.  
  The legends of these figures specify (1) $M_h$, the halo mass that the particle would be accreted onto in the absence of pressure, and (2) $M_J^{\rm max}$. 
We define $M_J^{\rm max}$ to be the maximum of the Jeans' mass evaluated at any snapshot before the particle crosses the $T_{\rm eq}$ curve.  The latter condition avoids including particles with high Jeans' masses that have been heated at the virial shock but  that still can cool and, hence, are not stabilized by pressure.  In the upper panels, particles with $M_h > M_J^{\rm max}$ are shown, while particles with $M_h < M_J^{\rm max}$ are shown in the lower panels.

Each simulated SPH particle's trajectory exhibits a vertical component at the reionization redshift.  After reionization, the typical particle evolves to lower densities and temperatures as the universe expands (and in accord with the picture presented in \S\ref{ss:example}). 
If the gas particle is being pulled into a halo with mass $M_h$ that is less massive than $M_J$ at any point during collapse (i.e., $<M_J^{\rm max}$), pressure is more likely to be able to overcome gravity and prevent collapse.  Hence, the particle will continue moving to lower densities in the Hubble flow if it has not reached turnaround before $M_h < M_J$ is satisfied.  
If the particle instead has passed turnaround once $M_h < M_J$ becomes satisfied, the trajectory likely will have decoupled from the Hubble flow when it is halted by pressure.  
Conversely, gas particles that satisfy $M_h > M_J$ over the entire trajectory likely collapse to high densities, radiating their gravitational energy and making it onto a galaxy.  However, one can see in Figures~\ref{fig:Tdsim}-\ref{fig:Tdsim3} that while these trends are present, sometimes this criterion errs, especially for particles for which $M_h \sim M_J^{\rm max}$.  We show later that a better criterion for collapse is $M_h > M_J^{\rm max}/4$ and even this revised condition for gravitational stability does not work for every particle.

The model presented in \S\ref{ss:example} assumed that heating from structure formation shocks is not important for determining whether gas can accrete.  The trajectories shown in Figures~\ref{fig:Tdsim}-\ref{fig:Tdsim3} add validation to this assumption.
Gas is sometimes heated at a halo's virial shock, causing it to reach temperatures significantly above the threshold for cooling (e.g., see the red dotted curve in the top panel of Fig.~3).  However, such gas cools down in a fraction of the age of the Universe for the $<10^{12}\Msun$ halos considered in this study and, hence, such heating does not impact whether the particle reaches high densities.  Gas can also shock at densities and temperatures where cooling is not efficient, and some of the trajectories in Fig.~\ref{fig:Tdsim} are moderately shock heated during this phase -- ascending to higher adiabats.  Such shocking becomes more probable with decreasing redshift and increasing halo mass.  However, a small fraction of the particles we follow at any redshift show significant shock heating during this phase, justifying our model's assumption.\footnote{Our simulations do not include galactic winds.  However, our picture shows that such winds would have the largest impact on suppressing accretion if they shock heated moderate overdensities that characterize the adiabatic collapse phase of the inflows.}

The overall tendency of the SPH particle trajectories shown in Figures~\ref{fig:Tdsim}-\ref{fig:Tdsim3} is consistent with the simple model illustrated in Figure~\ref{fig:theorycurve}.  Yet, there are two notable differences. 
First, the gas density at reionization varies by a factor of $\sim 3$ from our spherical collapse predictions (which are that the density is not much different than the cosmic mean density for the two lower redshift cases).  This variation owes to the gas not being in pressure equilibrium:  Before reionization heated the gas to $\sim 10^4$K, it clumped on much smaller scales than $k_J(T=10^4{\rm \;K})^{-1}$, and it then takes a while once being heated to thermally relax [a period of $\sim H(z)^{-1} (1+\delta_J)^{-1/2}$, where $\delta_J$ is the overdensity smoothed at the Jeans' scale].  It is the larger structures at the scale $\gtrsim k_J(T=10^4{\rm \;K})^{-1}$ that our model considers.  
Second, most gas particles first collapse into sheets and filaments (or reside in voids swept up by larger collapsing structures) before falling onto halos.  This results in turnaround (i.e. decoupling from the Hubble flow) occurring at different densities than in the spherical collapse model.  The Appendix discusses in more detail the applicability of spherical collapse for describing the density evolution of collapsing particles.

We have investigated how the trajectories are shaped by the amplitude of the ionizing background and the reionization redshift.  
We find that the characteristics of trajectories are not substantially altered if we use a $10$ times higher ionizing background.  In fact, the corresponding trajectories essentially trace each other in the simulation with $\Gamma_{-12} = 1$ and $\Gamma_{-12} = 10$ (SimG1 and SimG10):  Initially co-spatial particles follow the same path in $T-n_H$ until the particles reach densities at which cooling becomes important, which happens at somewhat higher densities in the case where $\Gamma_{-12} = 10$ than $\Gamma_{-12} = 1$.  Thus, the amplitude of the ionizing background has little impact on the halo mass that can accrete gas as $M_J^{\rm max}$ is only modestly increased if the SPH particle reaches somewhat higher densities before it can cool efficiently.  We also find that the redshift of reionization has little impact on accretion that occurs well after reionization.  In our picture, this results because both gas expanding and adiabatically collapsing encounters similar values of $(n_H, T)$ regardless of the redshift at which reionization happens.

\subsection{calibrating our simple picture}
\label{ss:simpic}

\begin{figure*}
\begin{center}
\includegraphics[width=14cm]{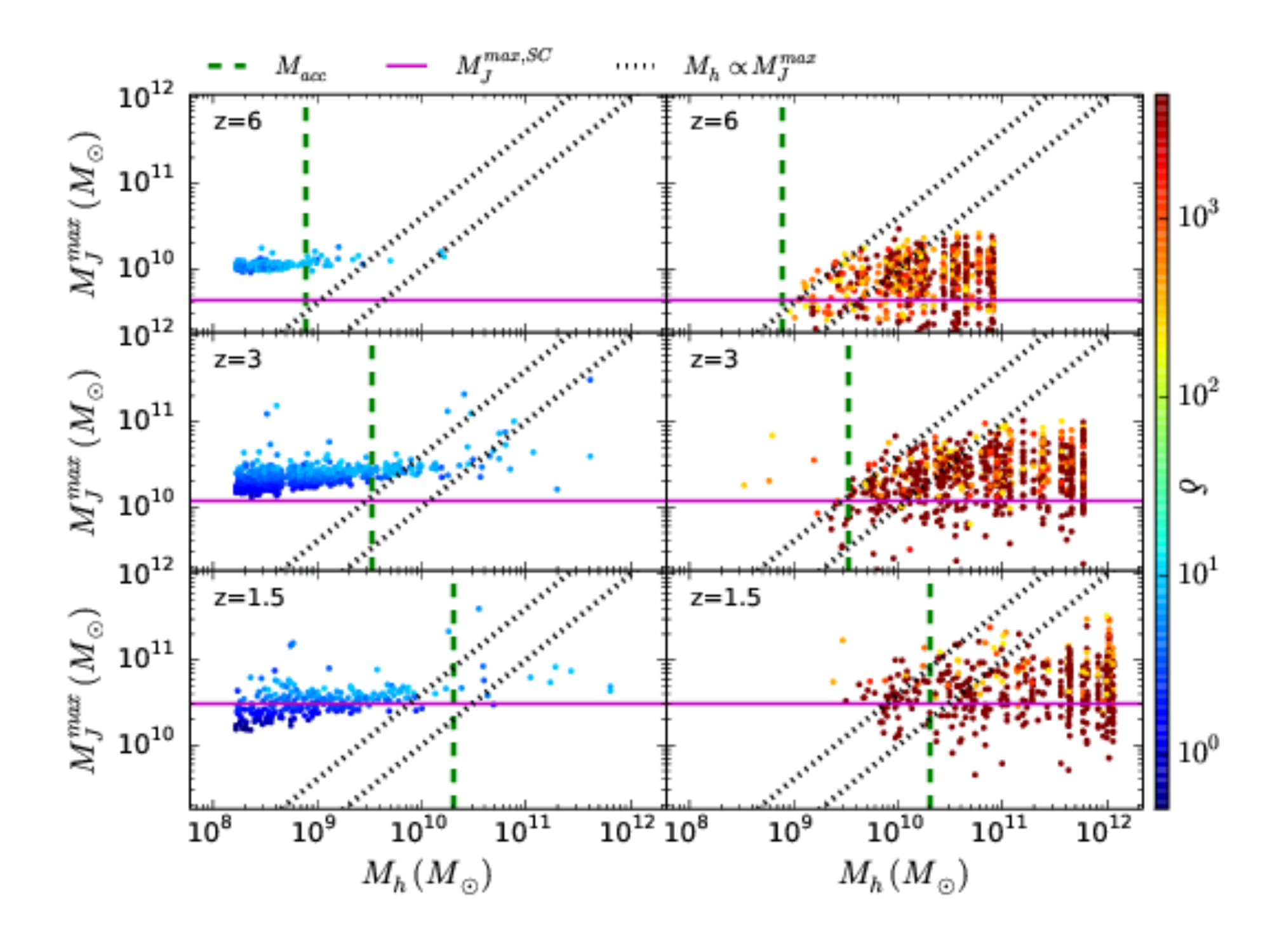}
\end{center}
\caption{The maximum Jeans' mass of gas particles prior to collapse ($z> \zcol$) in SimG1 as a function of the halo mass that they would be accreted onto in the absence of gas pressure. 
Each row of panels shows gas particles at $\zcol\approx 6,\ 3,$ and $1.5$,
from top to bottom. In each panel, the same number of gas particles are shown for each logarithmic halo mass bin.  The sampled particles correspond to those with $\delta_b< 10$ in the left panels and $\delta_b>200$ in the right.  The colors specify the densities, and SPH particles that have turned into star particles are given the maximum density on the colorbar. 
The two diagonal dotted lines are $M_h = M_J^{\rm max}/4$ and $M_h  = M_J^{\rm max}$, the horizontal solid line in each panel shows $M_J^{\rm max}$ if the trajectory follows spherical collapse, and the vertical dashed line is the accretion threshold of \citet{okamoto08}.
}
\label{fig:mGvsmJ}
\end{figure*}

Gas parcel which are Jeans' unstable over all density and temperatures they encounter tend to condense to high densities.  We argued that this amounts to the criterion that if the mass of the region is larger than $M_J^{\rm max}$, its gravity is sufficient to overcome pressure and gas is able to accrete, whereas if this is not satisfied, accretion is halted.  To test this criterion, Fig.~\ref{fig:mGvsmJ} plots $M_J^{\rm max}$ as a function of the halo mass at $z_{\rm coll}$ that the particle would collapse onto in the absence of pressure, making the ansatz that the halo mass at $z_{\rm coll}$ roughly approximates the ``mass of the region''.  In Figure~\ref{fig:mGvsmJ}, each row of panels shows gas particles accreted at $\zcol \approx 6,\ 3$ and $1.5$, from top to bottom.  
Where possible, the same number of gas particles are shown for each logarithmic halo mass bin.  The sampled particles are divided into different panels depending on their densities at $\zcol$.  In the lefthand panels, ``uncollapsed'' particles with $\delta_b< 10$ are shown, whereas the right shows ``collapsed'' particles with $\delta_b>200$, with the particle's color specifying the exact density.  Some of the $\delta_b>200$ particles have become star particles and are represented with the dark red color that indicates the color bar's maximum density.

In Figure~\ref{fig:mGvsmJ}, the two diagonal dotted lines show $M_h = M_J$ (bottom line) and $M_h =  M_J/4$ (top line).   The $M_{h} = M_{J}/4$ line does a good job at approximating the boundary between gas that cannot accrete onto halos (left panels) and that can (right panels), faring better than the $M_h = M_J$ demarkation.  That less massive halos can accrete than given by the criterion $M_h > M_J$ is not surprising.  The Jeans' length is the distance a sound wave can travel in a dynamical time.  However, after turnaround a gas parcel that is collapsing is in free-fall and adiabatically heating up, which means that a sound wave is not able to travel as far in Lagrangian space as one would predict from the instantaneous density and temperature of a gas particle.  Hence, density fluctuations are smoothed over a shorter distance than the Jeans' length, and the effective Jeans' mass is smaller, as we find.  However, this factor of $4$ tuning factor also reflects that the Jeans' mass does not provide the exact mass threshold that is able to accrete gas but rather a rough estimate for this mass.   Figure~\ref{fig:mGvsmJ} suggests that a tuning factor that is constant with redshift is sufficient.  This is not surprising as a spherically collapsing perturbation evolves through similar overdensities and temperatures independent of collapse redshift, at least well after reionization and prior to reaching densities and temperatures at which cooling is efficient.

The proportionality factor of $1/4$ also seems consistent with other indications of the fragmentation mass threshold above which cosmological clouds can collapse.  Studies of the relation between $N_{\rm HI}$ and $\delta_b$ in cosmological simulations find that models based on the Jeans' length predict $50\%$ larger densities at fixed the \HI\ column density $N_{\rm HI}$ \citep{mcquinn-LL, altay11}, which indicates that the Jeans' scale over-predicts the average size of overdense absorbers at a given density by $\sim50\%$ and, therefore, their mass by a factor of $\sim 1.5^3 = 3.4$. 

As mentioned above, the horizontal line in Figure~\ref{fig:mGvsmJ} shows the maximum Jeans mass if the trajectory follows spherical collapse and has $T=10^4~$K at turnaround. The systematic offset and dispersion around this horizontal line in the \emph{right} panels indicates that for most gas particles the collapse is not spherical.  Especially with decreasing redshift (where these halos are less rare), the collapse is first into sheets and then filaments.  In fact, interestingly we find that the average turnaround density, is somewhat lower than in spherical collapse, especially at $z=6$.  The Appendix investigates the spherical collapse approximation in more detail. 

Lastly, the long-dashed vertical lines in Figure~\ref{fig:mGvsmJ} show the instantaneous accretion prescription of \citet[eqn.~\ref{eqn:Ma}]{okamoto08}.  While this criterion differs from the diagonal curves that set the accretion prescription in our model, it still approximates reasonably well the transition mass between when gas is able to accrete and when it cannot.  This is likely why the \citet{okamoto08} prescription (applied on top of a merger tree) successful reproduces the cutoff mass found in simulations.

\section{Explaining the gas fraction of halos}
\label{sec:merger_tree}
\label{sec:fgas}

\begin{figure*}
\begin{center}
 \includegraphics[width=5.in]{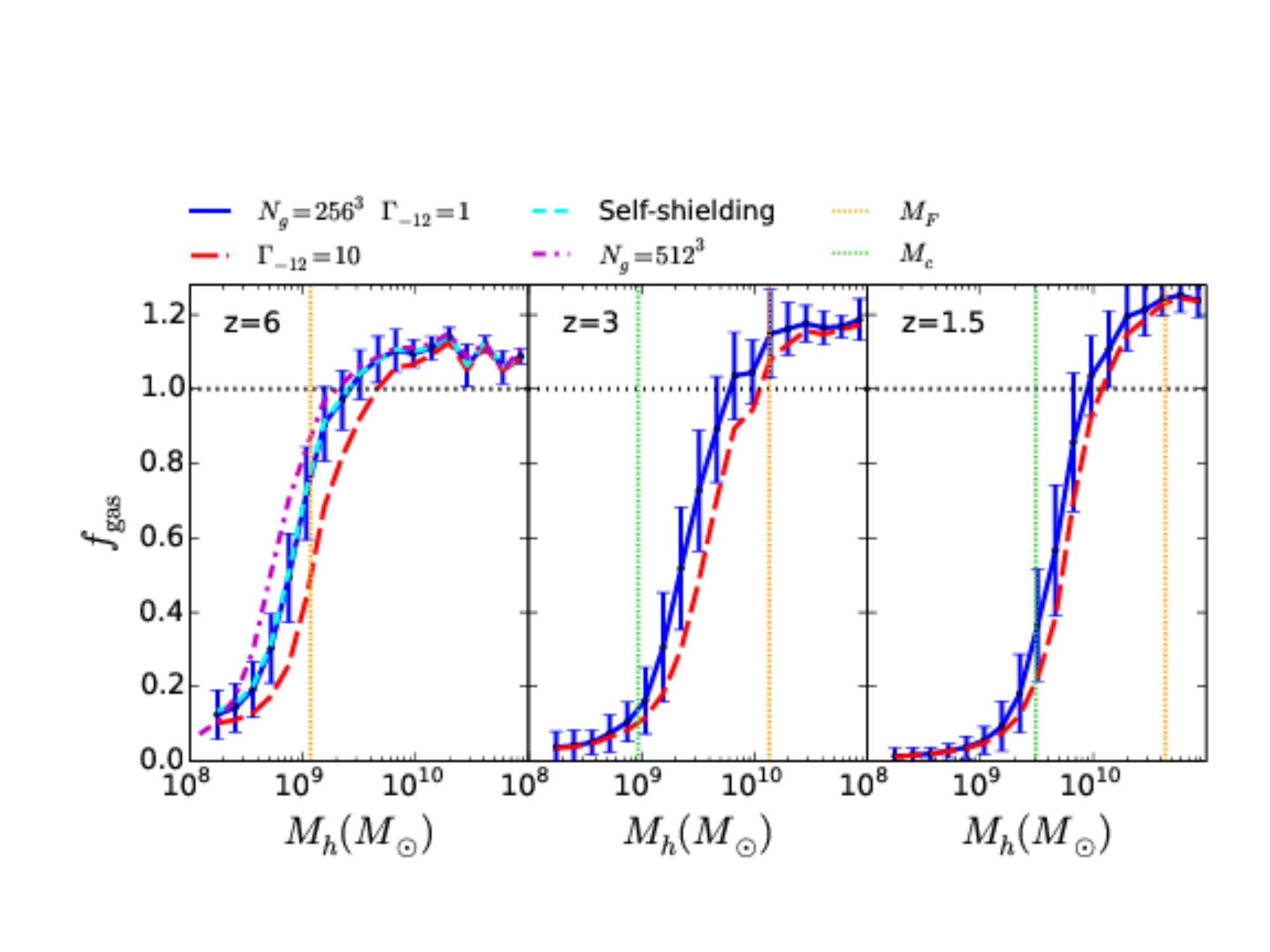}
 \end{center}
 \caption{Estimated gas fractions within one virial radius, $f_{\rm gas}$, as a function of halo mass.  The curves are computed by taking the ratio of the number of particles within a virial radius in the specified simulation to this number in the adiabatic simulation.  This procedure is done individually on each halo and the average is then taken in each halo mass bin.  Blue and red solid curves represent the gas fractions in the simulations with $\Gamma_{-12}=1$ and  $\Gamma_{-12}=10$ (SimG1 and SimG10), respectively. The error bars, shown only for the SimG1 case, give the standard deviation among the halos in each mass bin.
 The dotted horizontal lines show the maximum, $f_{\rm gas} = 1$, which the simulation estimates overshoot at the highest masses shown because the halos in the adiabatic simulation are puffier.  The cyan and purple curves in the leftmost, $z=6$ panel correspond to the simulations with self shielding (SimG1SS) and the $512^3$ gas particle simulation that has $8\times$ the resolution of SimG1 (SimG1N512), respectively.  The vertical long-dashed lines show the `characteristic mass' of \citet[leftmost line]{hoeft06} 
and the `filtering mass' of \citet[rightmost line]{gnedin00}.}
\label{fig:fgas}
\end{figure*}

Most studies of the impact of reionization on galaxies have concentrated on $f_{\rm gas}$ -- the gas mass fraction within a virial radius as a function of $M_h$ --, generally fitting for the halo mass that contains half of the baryons, $M_{1/2}$, as a function of redshift \citep{gnedin00, dijkstra04, hoeft06, sobacchi13a}.  Figure~\ref{fig:fgas} shows estimates of $f_{\rm gas}$ in our simulations at $z=1.5$, $3$, and $6$.   Our curves overshoot  $f_{\rm gas}=1$ at high masses because we calculate $f_{\rm gas}$ by dividing the number of gas particles found inside a virial radius in the specified simulation with the corresponding number in the adiabatic simulation:  The more massive gaseous halos in the adiabatic simulation tend to be puffier and, hence, have fewer particles within $r_{\rm vir}$.  Blue and red solid curves represent the gas fractions in the simulations with $\Gamma_{-12}=1$ and  $\Gamma_{-12}=10$ (SimG1 and SimG10), respectively. The error bars, shown only for the SimG1 case, give the standard deviation among the halos in each halo mass bin.  The cyan and purple curves in the leftmost, $z=6$ panel correspond to the simulations with self shielding (SimG1SS) and the simulation with $512^3$ gas particles (SimG1N512), respectively.    The small differences between these curves and those in the fiducial simulation, SimG1, illustrate that self-shielding has almost no effect and that the convergence in resolution is adequate.\footnote{The differences between the $256^3$ particle simulation and $512^3$ are largest at $z_{\rm coll}=6$ compared to the lower redshifts considered in the other panels in Figure~\ref{fig:fgas}.}

\begin{figure*}
\begin{center}
 \includegraphics[width=6.in]{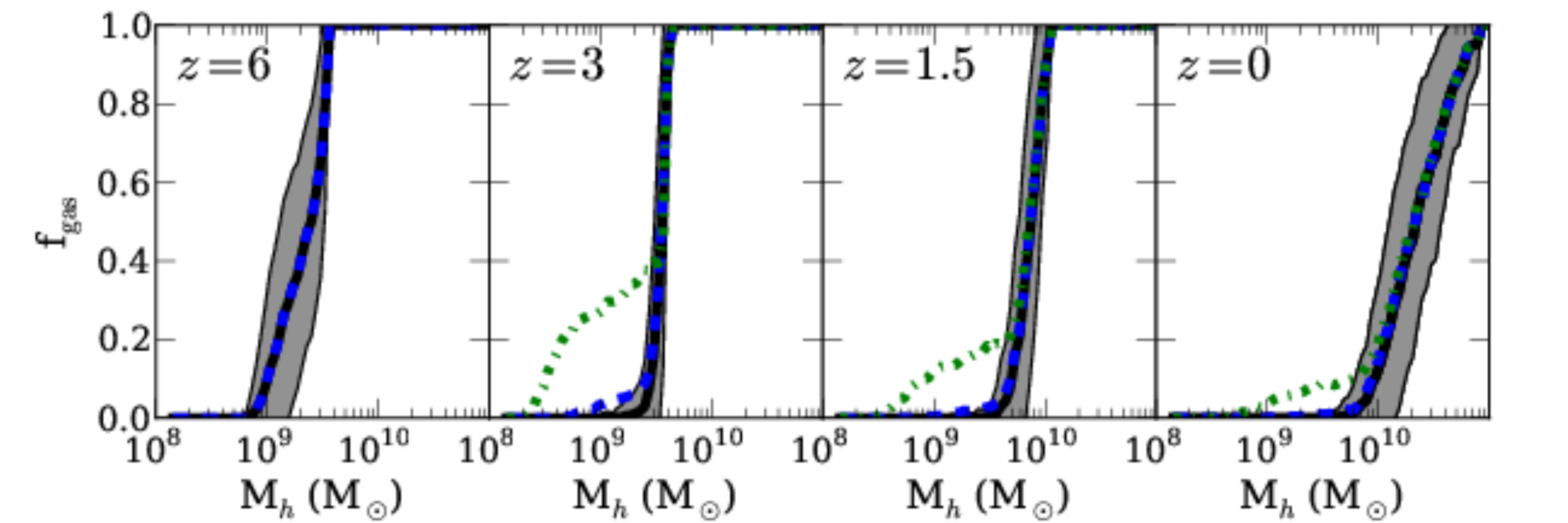}
 \end{center}
 \caption{The gas mass fraction, $f_{\rm gas}(M_h, z)$, in the merger tree--based model for accretion discussed in the text.  The black solid curves assume that halos below $3\times10^8~\Msun$ are photo-evaporated, and the grey highlighted regions show the standard deviation about the mean value of $f_{\rm gas}$.  The blue dashed curves are the case in which halos with masses below $1\times10^8~\Msun$ are photo-evaporated, and the green dot-dashed curves are the same but assume $z_{\rm rei} =6$ rather than the fiducial value of $z_{\rm rei} =9$.  The curves show a similar $M_{1/2}$ compared to the simulations and also have a similar standard deviation in $f_{\rm gas}$ (compare with Fig.~\ref{fig:fgas}).  However, the transition from zero to one is more abrupt in the merger tree calculations than in the simulations (particularly at $z=3$).  
 \label{fig:fgasSC}}
\end{figure*}

 The $f_{\rm gas}$ estimates in Figure~\ref{fig:fgas} illustrate the result that there is a relatively well defined halo mass at each redshift that can accrete gas, with the transition from almost no gas to the cosmic closure density of gas occurring over only a factor of $\sim  3$ in mass.  There is a weak increase in $M_{1/2}$ when increasing $\Gamma_{-12}$ (compare the blue and red curves which correspond to $\Gamma_{-12}=1$ and  $\Gamma_{-12}=10$), with the most significant difference occurring in the $z=6$ panel.  This is consistent with our picture in which an increase in $\Gamma_{-12}$ has a more prominent (but still small) effect at higher redshifts (see Fig.~\ref{fig:theorycurve}).

 The vertical dotted lines show the halo masses that contain half of their gas, $M_{1/2}$, in the models of \citet[eqn.~\ref{eqn:Mc}]{hoeft06} and \citet[eqn.~\ref{eqn:Mf}]{gnedin00}.   We do not show the mass threshold in \citet{okamoto08} as their prescription (like ours) predicts the instantaneous accretion mass and not $M_{1/2}$.   Both the \citet{hoeft06}and \citet{gnedin00} mass thresholds do not match the $M_{1/2}$ found in the simulations.

To test whether our simple prescription for instantaneous accretion explains the simulations' $f_{\rm gas}(M_h)$, we implement the prescription described in the previous section on top of a halo merger tree calculation, using the \citet{neistein08} merger tree code and assuming spherical collapse.  Specifically, we assumed that gas with turnaround redshift after reionization is at a temperature of $10^4~$K until its turnaround redshift in spherical collapse, at which time it is heated adiabatically as it collapses.  
We define $M_J^{\rm max, SC}(z)$ to be the Jeans' mass at the point this adiabatic trajectory intersects the equilibrium temperature, where $z$ denotes the collapse redshift.   In the merger tree, if $M_h(z) > M_J^{\rm max, SC}(z)/4$ is satisfied for a halo of mass $M_h$ then the gas can be accreted, where $1/4$ is the calibration factor that we found in the previous section.  If this criterion is not satisfied, gas is not accreted until a redshift at which the halo has acquired enough dark matter that the criterion \emph{is} satisfied.  At that point, the halo can accrete all of the gas that it had previously been unable to accrete.  For gas that has a turnaround redshift before reionization but collapse redshift after reionization, we instead apply the criterion for accretion $M_h(z) > M_J(n_{\rm rei}, T_{\rm eq}(n_{\rm rei}))/4$, where $n_{\rm rei}$ is the hydrogen number density of the spherically collapsing parcel at reionization. 
  
The merger tree calculations also must treat halos that formed prior to reionization.  Such halos were able to pull in gas down to much lower masses, although many of these halos would also have been photo-evaporated prior to merging into larger systems.  \citet{barkana04} found with 1D radiative transfer calculations that $M_h \lesssim 10^8\Msun$ halos are photo-evaporated by the ionizing background, and they found that this mass is only modestly affected by self-shielding.\footnote{Our self-shielding criterion (eqn.~\ref{eqn:schayer0}) would imply that bound gas within $r_{200}$ of \emph{any} halo would be able to fully self-shield to the ionizing background at $z\gtrsim8$ for $\Gamma_{-12} = 0.1$ and $T=10^4~$K.  However, this criterion assumes that the scale of fluctuations is the Jeans' scale for $10^4~$K gas, which does not apply to unheated gas or gas that is relaxing after recently being heated (and hence to the photo-evaporation of halos).}  Our merger tree algorithm leaves the exact mass scale below which halos can be evaporated as a free parameter, which we set to $1\times10^8\Msun$ and $3\times10^8\Msun$.  The former is closer to the results of \citet{barkana04} and the latter to the halo mass ``minimally'' resolved with $50$ particles in our $256^3$ simulation.

Figure~\ref{fig:fgasSC} shows the results of this merger tree calculation for $\Gamma_{-12} = 1$ and at four redshifts.  The black solid curves take halos with masses below $1\times10^8~\Msun$ to be photo-evaporated, whereas the blue dashed curves take this mass to be $3\times10^8~\Msun$.  The differences between these two cases are modest, being largest in the $z=6$ panel. This figure should be compared to the results from the simulations (Fig.~\ref{fig:fgas}).  The leftmost three panels in Figure~\ref{fig:fgasSC} show the same $z_{\rm coll}$ as the three panels in Figure~\ref{fig:fgas}.  The two figures by-in-large show agreement both in $M_{1/2}$ and the standard deviation in $f_{\rm gas}(M_h)$.  However, especially at intermediate redshifts, the merger tree $f_{\rm gas}(M_h)$ shows a more abrupt transition from zero to one than the $f_{\rm gas}(M_h)$ in the simulations.  This difference likely owes to the merger tree's simplification of trajectories following spherical collapse and hence having a more uniform $M_J^{\rm max}$ (see the appendix).  Still, by $z=0$ there is a large range of masses in the merger calculations that have $0 < f_{\rm gas}<1$.

All of our calculations to this point have taken $z_{\rm rei}=9$.  In reality, reionization is a complicated process that should span a significant duration in redshift.  Some regions may have been reionized as late as $z_{\rm rei}=6$, whereas half of the gas was likely ionized at $z>10$ \citep{iliev06, mcquinn-morphology, trac07, busha10,lunnan12}.  The dot-dashed green curves in Figure~\ref{fig:fgasSC} show the results of the merger tree for $z_{\rm rei}=6$.  In this case, more of a tail develops to low halo masses in the mean $f_{\rm gas}$.  While not shown in Figure~\ref{fig:fgasSC}, the standard deviation in $f_{\rm gas}$ for the  $z_{\rm rei}=6$ case is comparable to the mean.  For the Milky Way's ultra-faint dwarfs, the characteristics of this tail (which relates to when the Local Group was reionized) likely influence their nature.  This also implies that the properties of dwarf galaxies varies throughout the Universe depending on the redshift of reionization \citep{milosavljevic13}.

\section{Conclusions}

This paper developed and tested with simulations an intuitive model for how the interplay between gravity, pressure, cooling, and self-shielding set the redshift--dependent mass scale at which halos can accrete gas.  This model is based on how the evolution of a collapsing gas cloud is bounded by several critical curves in density--temperature space.  This model explains why gas accretion onto halos well after reionization is neither strongly impacted by the amplitude of the ionizing background nor the reionization redshift.\footnote{The lack of dependence on $\Gamma_{-12}$ rules out suggestions that this dependence could regulate star formation and explain why the Universe has what appears to be an extremely-fine tuned, nearly constant ionizing background over $z=2-5$ \citep{faucher08, mcquinn-LL}.}

Previous analytic and semi-analytic models assumed that the halo mass threshold above which halos can pull in surrounding gas corresponds to the Jeans' mass or its cosmological analog, the filtering mass.  To determine whether a halo should contain gas, these mass scales had either been evaluated at the mean density of the Universe \citep{shapiro94, gnedin00, busha10, lunnan12} or at densities near the halo virial density \citep{ hoeft06, okamoto08}, leading to a factor of $\sim 10$ difference in the predicted mass threshold.  We showed that neither of these prescriptions is quite right:  A spherically collapsing gas parcel encounters densities that are never within an order of magnitude of the cosmic mean density at its collapse redshift.  
Furthermore, by the time it encounters densities comparable to the virial density of a halo, it will almost certainly continue collapsing (and ultimately accrete onto a galaxy) as it is able to radiate away its energy efficiently.  The bottleneck for collapse occurs at densities that are an order of magnitude lower than the virial density and an order of magnitude higher than the cosmic mean density, densities at which gas is not yet able to cool efficiently.

Our model depends on the formation history of a halo rather than its instantaneous halo mass.  Capturing the formation history is critical for (1) exploring the hypothesis that the ultra-faint dwarfs formed prior to reionization, and (2) understanding which halos are accreting gas at a given redshift (and hence likely to be forming stars).    
Regarding (1), our model enabled us to calculate how $f_{\rm gas}$ depends on the local reionization redshift, $z_{\rm rei}$.  While it predicts that the halo mass $M_h$ at which $f_{\rm gas}(M_h)=0.5$ is unaffected by $z_{\rm rei}$, our model finds that the number of halos with $f_{\rm gas}(M_h)\lesssim0.2$ is very sensitive to $z_{\rm rei}$.  Regarding (2), at $z=0$ even a $10^{11}\Msun$ halo is unable to accrete unshocked intergalactic gas, which is a factor of several larger than the mass scale at which $f_{\rm gas}(M_h) = 0.5$.  

A significant drawback of our model is that it lacks a precise analytic criterion for gravitational instability and instead assumes that the masses that are gravitationally  unstable are given by a redshift-independent constant times the maximum Jeans' mass a gas parcel obtains during collapse.  We calibrated this constant with simulations, and gave a physical motivation for the derived value.  Once this calibration factor was determined, our model was able to reproduce the mass scale at which $f_{\rm gas}(M_h) = 0.5$ when implemented on top of a halo merger tree calculation.

Our study ignored the impact of outflows from stellar feedback.  Outflows may terminate the inflows studied here, preventing them from fueling galaxies.  However, even if outflows have a large effect on subsequent accretion, our calculations still determine which halos were able to accrete gas, form stars, and hence later have outflows. \\

We thank Andrei Mesinger, Robert Feldman, Nick Gnedin, and Smadar Naoz for useful discussions.  In addition, we thank our referee Takashi Okamoto for comments that improved the manuscript.  MM acknowledges support by the National Aeronautics and Space Administration through the Hubble Postdoctoral Fellowship and also from NSF grant AST~1312724.

\bibliographystyle{apj}
\bibliography{References}

\begin{thebibliography}{}

\bibitem[\protect\citeauthoryear{{Altay} et~al.}{{Altay}
  et~al.}{2011}]{altay11}
{Altay}, G., {Theuns}, T., {Schaye}, J., {Crighton}, N.~H.~M.,  \& {Dalla
  Vecchia}, C. 2011, \apjl, 737, L37

\bibitem[\protect\citeauthoryear{{Barkana} \& {Loeb}}{{Barkana} \&
  {Loeb}}{2004}]{barkana04}
{Barkana}, R.,  \& {Loeb}, A. 2004, \apj, 609, 474

\bibitem[\protect\citeauthoryear{{Becker} \& {Bolton}}{{Becker} \&
  {Bolton}}{2013}]{becker13}
{Becker}, G.~D.,  \& {Bolton}, J.~S. 2013, ArXiv:1307.2259

\bibitem[\protect\citeauthoryear{{Benson} et~al.}{{Benson}
  et~al.}{2002}]{benson02}
{Benson}, A.~J., {Frenk}, C.~S., {Lacey}, C.~G., {Baugh}, C.~M.,  \& {Cole}, S.
  2002, \mnras, 333, 177

\bibitem[\protect\citeauthoryear{{Binney} \& {Tremaine}}{{Binney} \&
  {Tremaine}}{1987}]{binney87}
 1987, {Galactic dynamics}, ed. {Binney, J.~\& Tremaine, S.}

\bibitem[\protect\citeauthoryear{{Bolton} et~al.}{{Bolton}
  et~al.}{2005}]{bolton05}
{Bolton}, J.~S., {Haehnelt}, M.~G., {Viel}, M.,  \& {Springel}, V. 2005,
  \mnras, 357, 1178

\bibitem[\protect\citeauthoryear{{Brown} et~al.}{{Brown}
  et~al.}{2012}]{brown12}
{Brown}, T.~M., et~al. 2012, \apjl, 753, L21

\bibitem[\protect\citeauthoryear{{Bullock}, {Kravtsov}, \&
  {Weinberg}}{{Bullock} et~al.}{2000}]{bullock00}
{Bullock}, J.~S., {Kravtsov}, A.~V.,  \& {Weinberg}, D.~H. 2000, \apj, 539, 517

\bibitem[\protect\citeauthoryear{{Busha} et~al.}{{Busha}
  et~al.}{2010}]{busha10}
{Busha}, M.~T., {Alvarez}, M.~A., {Wechsler}, R.~H., {Abel}, T.,  \&
  {Strigari}, L.~E. 2010, \apj, 710, 408

\bibitem[\protect\citeauthoryear{{Calverley} et~al.}{{Calverley}
  et~al.}{2011}]{calverley11}
{Calverley}, A.~P., {Becker}, G.~D., {Haehnelt}, M.~G.,  \& {Bolton}, J.~S.
  2011, \mnras, 412, 2543

\bibitem[\protect\citeauthoryear{{Crocce}, {Pueblas}, \&
  {Scoccimarro}}{{Crocce} et~al.}{2006}]{crocce06}
{Crocce}, M., {Pueblas}, S.,  \& {Scoccimarro}, R. 2006, \mnras, 373, 369

\bibitem[\protect\citeauthoryear{{Dekel} \& {Woo}}{{Dekel} \&
  {Woo}}{2003}]{dekel03}
{Dekel}, A.,  \& {Woo}, J. 2003, \mnras, 344, 1131

\bibitem[\protect\citeauthoryear{{Dijkstra} et~al.}{{Dijkstra}
  et~al.}{2004}]{dijkstra04}
{Dijkstra}, M., {Haiman}, Z., {Rees}, M.~J.,  \& {Weinberg}, D.~H. 2004, \apj,
  601, 666

\bibitem[\protect\citeauthoryear{{Fan} et~al.}{{Fan} et~al.}{2006}]{fan06}
{Fan}, X., et~al. 2006, \aj, 132, 117

\bibitem[\protect\citeauthoryear{{Faucher-Gigu{\`e}re}
  et~al.}{{Faucher-Gigu{\`e}re} et~al.}{2010}]{faucher10}
{Faucher-Gigu{\`e}re}, C.-A., {Kere{\v s}}, D., {Dijkstra}, M., {Hernquist},
  L.,  \& {Zaldarriaga}, M. 2010, \apj, 725, 633

\bibitem[\protect\citeauthoryear{{Faucher-Gigu{\`e}re}
  et~al.}{{Faucher-Gigu{\`e}re} et~al.}{2008}]{faucher08}
{Faucher-Gigu{\`e}re}, C.-A., {Lidz}, A., {Hernquist}, L.,  \& {Zaldarriaga},
  M. 2008, \apj, 688, 85

\bibitem[\protect\citeauthoryear{{Field}}{{Field}}{1965}]{field65}
{Field}, G.~B. 1965, \apj, 142, 531

\bibitem[\protect\citeauthoryear{{Finlator}, {Dav{\'e}}, \&
  {{\"O}zel}}{{Finlator} et~al.}{2011}]{finlator11}
{Finlator}, K., {Dav{\'e}}, R.,  \& {{\"O}zel}, F. 2011, \apj, 743, 169

\bibitem[\protect\citeauthoryear{{Gnedin}}{{Gnedin}}{2000}]{gnedin00}
{Gnedin}, N.~Y. 2000, \apj, 542, 535

\bibitem[\protect\citeauthoryear{{Gnedin} \& {Hui}}{{Gnedin} \&
  {Hui}}{1998}]{gnedin98}
{Gnedin}, N.~Y.,  \& {Hui}, L. 1998, \mnras, 296, 44

\bibitem[\protect\citeauthoryear{{Gunn} \& {Gott}}{{Gunn} \&
  {Gott}}{1972}]{gunn72}
{Gunn}, J.~E.,  \& {Gott}, J.~R., III. 1972, \apj, 176, 1

\bibitem[\protect\citeauthoryear{{Haardt} \& {Madau}}{{Haardt} \&
  {Madau}}{1996}]{haardt96}
{Haardt}, F.,  \& {Madau}, P. 1996, \apj, 461, 20

\bibitem[\protect\citeauthoryear{{Haardt} \& {Madau}}{{Haardt} \&
  {Madau}}{2012}]{haardt12}
{Haardt}, F.,  \& {Madau}, P. 2012, \apj, 746, 125

\bibitem[\protect\citeauthoryear{{Hoeft} et~al.}{{Hoeft}
  et~al.}{2006}]{hoeft06}
{Hoeft}, M., {Yepes}, G., {Gottl{\"o}ber}, S.,  \& {Springel}, V. 2006, \mnras,
  371, 401

\bibitem[\protect\citeauthoryear{{Hui} \& {Gnedin}}{{Hui} \&
  {Gnedin}}{1997}]{hui97}
{Hui}, L.,  \& {Gnedin}, N.~Y. 1997, \mnras, 292, 27

\bibitem[\protect\citeauthoryear{{Iliev} et~al.}{{Iliev}
  et~al.}{2006}]{iliev06}
{Iliev}, I.~T., {Mellema}, G., {Pen}, U., {Merz}, H., {Shapiro}, P.~R.,  \&
  {Alvarez}, M.~A. 2006, \mnras, 369, 1625

\bibitem[\protect\citeauthoryear{{Kere{\v s}} et~al.}{{Kere{\v s}}
  et~al.}{2005}]{keres05}
{Kere{\v s}}, D., {Katz}, N., {Weinberg}, D.~H.,  \& {Dav{\'e}}, R. 2005,
  \mnras, 363, 2

\bibitem[\protect\citeauthoryear{{Larson} et~al.}{{Larson}
  et~al.}{2011}]{larson11}
{Larson}, D.,  et~al. 2011, \apjs, 192, 16

\bibitem[\protect\citeauthoryear{{Lunnan} et~al.}{{Lunnan}
  et~al.}{2012}]{lunnan12}
{Lunnan}, R., {Vogelsberger}, M., {Frebel}, A., {Hernquist}, L., {Lidz}, A.,
  \& {Boylan-Kolchin}, M. 2012, \apj, 746, 109

\bibitem[\protect\citeauthoryear{{Mashchenko}, {Wadsley}, \&
  {Couchman}}{{Mashchenko} et~al.}{2008}]{mashchenko08}
{Mashchenko}, S., {Wadsley}, J.,  \& {Couchman}, H.~M.~P. 2008, Science, 319,
  174

\bibitem[\protect\citeauthoryear{{McQuinn}}{{McQuinn}}{2012}]{mcquinn-Xray}
{McQuinn}, M. 2012, \mnras, 426, 1349

\bibitem[\protect\citeauthoryear{{McQuinn} et~al.}{{McQuinn}
  et~al.}{2007}]{mcquinn-morphology}
{McQuinn}, M., {Lidz}, A., {Zahn}, O., {Dutta}, S., {Hernquist}, L.,  \&
  {Zaldarriaga}, M. 2007, \mnras, 377, 1043

\bibitem[\protect\citeauthoryear{{McQuinn} et~al.}{{McQuinn}
  et~al.}{2009}]{mcquinn-HeII}
{McQuinn}, M., {Lidz}, A., {Zaldarriaga}, M., {Hernquist}, L., {Hopkins},
  P.~F., {Dutta}, S.,  \& {Faucher-Gigu{\`e}re}, C.-A. 2009, \apj, 694, 842

\bibitem[\protect\citeauthoryear{McQuinn, Oh, \& Faucher-Giguere}{McQuinn
  et~al.}{2011}]{mcquinn-LL}
McQuinn, M., Oh, S.,  \& Faucher-Giguere, C.-A. 2011, Astrophys.J., 743, 82

\bibitem[\protect\citeauthoryear{{McQuinn} \& {Worseck}}{{McQuinn} \&
  {Worseck}}{2013}]{mcquinn13}
{McQuinn}, M.,  \& {Worseck}, G. 2013, ArXiv:1306.4985

\bibitem[\protect\citeauthoryear{{Milosavljevic} \& {Bromm}}{{Milosavljevic} \&
  {Bromm}}{2013}]{milosavljevic13}
{Milosavljevic}, M.,  \& {Bromm}, V. 2013, ArXiv e-prints

\bibitem[\protect\citeauthoryear{{Miralda-Escud{\'e}}}{{Miralda-Escud{\'e}}}{2005}]{miralda05}
{Miralda-Escud{\'e}}, J. 2005, \apjl, 620, L91

\bibitem[\protect\citeauthoryear{{Miralda-Escud{\'e}} \&
  {Rees}}{{Miralda-Escud{\'e}} \& {Rees}}{1994}]{miralda94}
{Miralda-Escud{\'e}}, J.,  \& {Rees}, M.~J. 1994, \mnras, 266, 343

\bibitem[\protect\citeauthoryear{{Naoz}, {Barkana}, \& {Mesinger}}{{Naoz}
  et~al.}{2009}]{2009MNRAS.399..369N}
{Naoz}, S., {Barkana}, R.,  \& {Mesinger}, A. 2009, \mnras, 399, 369

\bibitem[\protect\citeauthoryear{{Neistein} \& {Dekel}}{{Neistein} \&
  {Dekel}}{2008}]{neistein08}
{Neistein}, E.,  \& {Dekel}, A. 2008, \mnras, 383, 615

\bibitem[\protect\citeauthoryear{{Okamoto} \& {Frenk}}{{Okamoto} \&
  {Frenk}}{2009}]{okamoto09}
{Okamoto}, T.,  \& {Frenk}, C.~S. 2009, \mnras, 399, L174

\bibitem[\protect\citeauthoryear{{Okamoto} et~al.}{{Okamoto}
  et~al.}{2010}]{okamoto10}
{Okamoto}, T., {Frenk}, C.~S., {Jenkins}, A.,  \& {Theuns}, T. 2010, \mnras,
  406, 208

\bibitem[\protect\citeauthoryear{{Okamoto}, {Gao}, \& {Theuns}}{{Okamoto}
  et~al.}{2008}]{okamoto08}
{Okamoto}, T., {Gao}, L.,  \& {Theuns}, T. 2008, \mnras, 390, 920

\bibitem[\protect\citeauthoryear{{Pawlik} \& {Schaye}}{{Pawlik} \&
  {Schaye}}{2009}]{pawlik09b}
{Pawlik}, A.~H.,  \& {Schaye}, J. 2009, \mnras, 396, L46

\bibitem[\protect\citeauthoryear{{Pe{\~n}arrubia}, {Navarro}, \&
  {McConnachie}}{{Pe{\~n}arrubia} et~al.}{2008}]{penarrubia08}
{Pe{\~n}arrubia}, J., {Navarro}, J.~F.,  \& {McConnachie}, A.~W. 2008, \apj,
  673, 226

\bibitem[\protect\citeauthoryear{{Pontzen} \& {Governato}}{{Pontzen} \&
  {Governato}}{2012}]{pontzen12}
{Pontzen}, A.,  \& {Governato}, F. 2012, \mnras, 421, 3464

\bibitem[\protect\citeauthoryear{{Quinn}, {Katz}, \& {Efstathiou}}{{Quinn}
  et~al.}{1996}]{quinn96}
{Quinn}, T., {Katz}, N.,  \& {Efstathiou}, G. 1996, \mnras, 278, L49

\bibitem[\protect\citeauthoryear{{Rahmati} et~al.}{{Rahmati}
  et~al.}{2013}]{rahmati13}
{Rahmati}, A., {Schaye}, J., {Pawlik}, A.~H.,  \& {Raicevic}, M. 2013, \mnras,
  431, 2261

\bibitem[\protect\citeauthoryear{{Schaye}}{{Schaye}}{2001}]{schaye01}
{Schaye}, J. 2001, \apj, 559, 507

\bibitem[\protect\citeauthoryear{{Shapiro}, {Giroux}, \& {Babul}}{{Shapiro}
  et~al.}{1994}]{shapiro94}
{Shapiro}, P.~R., {Giroux}, M.~L.,  \& {Babul}, A. 1994, \apj, 427, 25

\bibitem[\protect\citeauthoryear{{Sobacchi} \& {Mesinger}}{{Sobacchi} \&
  {Mesinger}}{2013}]{sobacchi13a}
{Sobacchi}, E.,  \& {Mesinger}, A. 2013, \mnras, 432, L51

\bibitem[\protect\citeauthoryear{{Somerville}}{{Somerville}}{2002}]{somerville02}
{Somerville}, R.~S. 2002, \apjl, 572, L23

\bibitem[\protect\citeauthoryear{{Springel}, {Yoshida}, \& {White}}{{Springel}
  et~al.}{2001}]{springel01}
{Springel}, V., {Yoshida}, N.,  \& {White}, S.~D.~M. 2001, New~Astronomy, 6, 79

\bibitem[\protect\citeauthoryear{{Thoul} \& {Weinberg}}{{Thoul} \&
  {Weinberg}}{1996}]{thoul96}
{Thoul}, A.~A.,  \& {Weinberg}, D.~H. 1996, \apj, 465, 608

\bibitem[\protect\citeauthoryear{{Trac} \& {Cen}}{{Trac} \&
  {Cen}}{2007}]{trac07}
{Trac}, H.,  \& {Cen}, R. 2007, \apj, 671, 1

\bibitem[\protect\citeauthoryear{{Worseck} et~al.}{{Worseck}
  et~al.}{2011}]{worseck11}
{Worseck}, G., et~al. 2011, \apjl, 733, L24

\end{thebibliography}

\appendix

\section{trajectories of gas particles in $n_H-T$ plane}

\begin{figure*}
\begin{center}
 \includegraphics[width=2.3in]{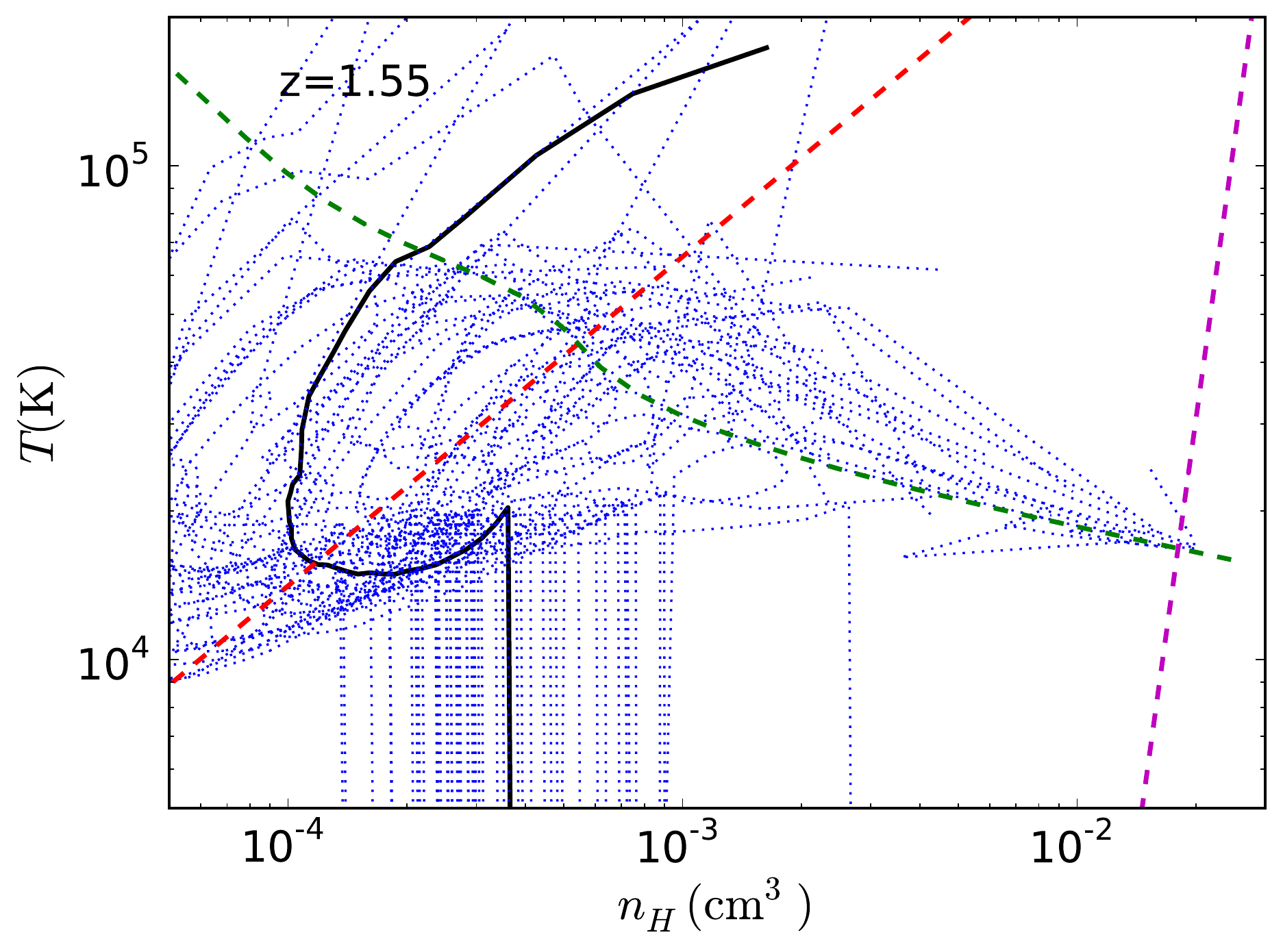} \includegraphics[width=2.3in]{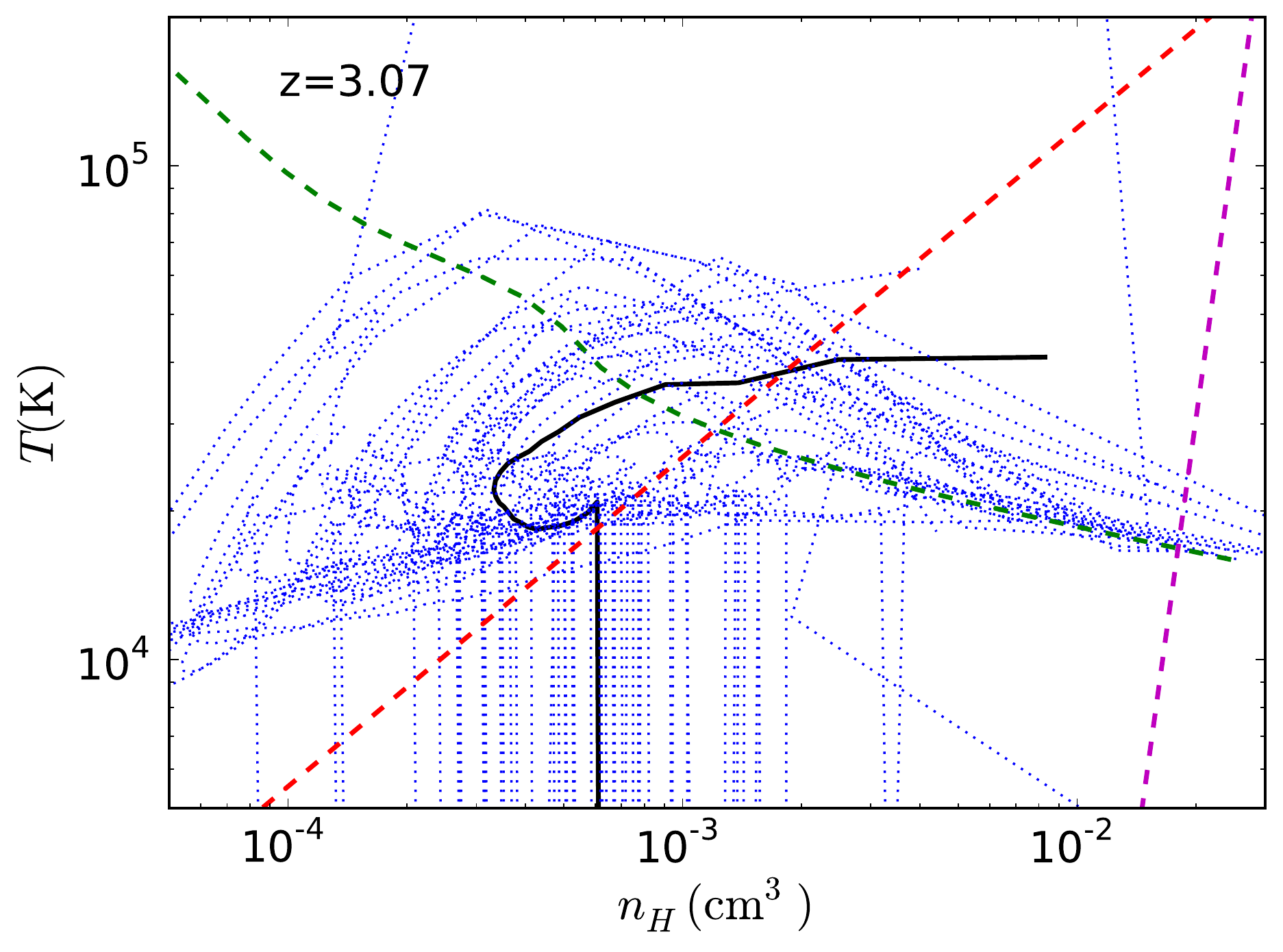} \includegraphics[width=2.3in]{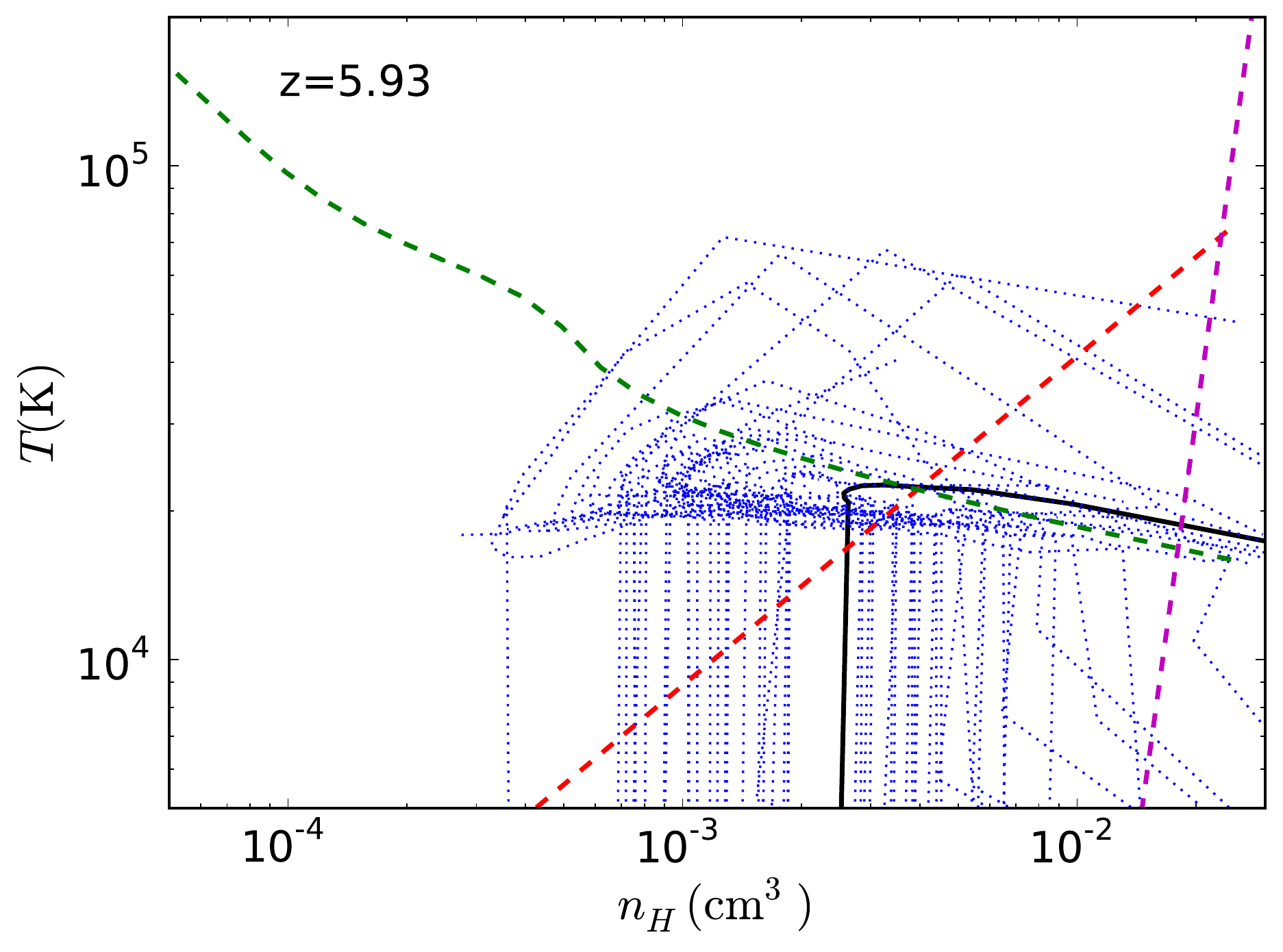}
\end{center}
   \caption{The trajectories of particles that collapsed at $z=1.5$ (left panel), $z=3$ (middle panel), and $z=6$ (right panel). The black thick solid curves represent the average trajectory of the particles that crossed the overdensity threshold of $\delta_b > 200$ in SimG1 at the specified redshift.  The blue dotted curves show the trajectories of fifty gas particles randomly selected from among the particles used in the average.  The three dashed curves are included for reference and show $T_{\rm eq}$ (green curves), the adiabat for gas at $T=10^{4}$ at turnaround in the spherical collapse model ($\delta_{\rm ta} = 4.6$, red curves), and the self-shielding threshold given by equation~\ref{eqn:SS}  (magenta curves).  This figure illustrates the approximate nature of the spherical collapse model, which predicts a single $n_H-T$ trajectory if the temperature at turnaround is held fixed:  The trajectories of the particles that collapse at a single redshift go through a wide range of temperatures and densities.
\label{fig:Tdave1}}
\end{figure*}

In Figure~\ref{fig:Tdave1}, the black thick solid curves show the average trajectories of SimG1 gas particles selected to have $\delta_b > 200$ at $\zcol$ for $\zcol = 6,~ 3,$ and $1.5$.
The particles used in this average were selected from the subset of gas particles that would have been accreted in SimAd using the accretion criterion described in \S4.
The average is taken in $\log n_H-\log T$ space. 
In each figure, fifty trajectories of gas particles (which were randomly selected from among the particles that were used to compute the average trajectory) are shown with the blue dotted lines.  In addition, the green dashed curve is the equilibrium temperature at which photoheating balances atomic cooling. 
The red dashed line is the adiabat that the gas particles would follow if they collapsed at the specified redshift following spherical collapse, with $T=10^4$ at turnaround. 
The nearly vertical magenta dashed line is the threshold at which gas fully self-shields to hydrogen photoionizations (eqn.~\ref{eqn:SS}).  

The fifty trajectories show that the gas particles tend to turnaround at somewhat lower density than the turnaround density of $5.6 \langle n_H \rangle$ in the spherical collapse model.  However, the main trend that is illustrated is that there is broad dispersion in the turnaround redshift.  This trend was also apparent in Figure~\ref{fig:mGvsmJ} (especially with increasing $z$), as the $M_J^{\rm max}$ of particles in the simulations was likely to be larger than $M_J^{\rm max, SC}$ estimated using spherical collapse model.  In addition, the average temperature of the trajectories that collapse tends to lie above $T_{\rm eq}$ at the higher densities shown owing to shock heating at virialization.

\end{document}